\documentclass[final]{siamltex}  % Use this line for a4

\usepackage{amssymb}
\usepackage{amsxtra}

\newtheorem{thm}{Theorem}[section]
\newtheorem{remark}{Remark}[section]
\newtheorem{cor}{Corollary}[section]

\numberwithin{equation}{section}

\newcommand{\F}{\mathcal{F}}
\newcommand{\G}{\mathcal{G}}
\newcommand{\Fb}{\mathbb{F}}
\newcommand{\Gb}{\mathbb{G}}
\renewcommand{\P}{\mathbb{P}}
\newcommand{\E}{\mathbb{E}}
\newcommand{\m}{\Phi^{\times}}
\newcommand{\p}{\Phi^{+}}
\newcommand{\ms}{\phi^{\times}}
\newcommand{\ps}{\phi^{+}}
\newcommand{\eps}{\varepsilon}

\title{\LARGE \bf
\sc Quickest Detection of a Minimum of Two Poisson Disorder Times}

\author{Erhan Bayraktar % <-this % stops a space
\thanks{E. Bayraktar is with the Department of Mathematics, University
of Michigan, Ann Arbor, MI 48109, USA, email: erhan@umich.edu} \and  H. Vincent Poor%
\thanks{H. V. Poor is with the School of Engineering and Applied Science, Princeton University,
        Princeton, NJ 08544, USA, email: poor@princeton.edu}%
}
\date{}

\begin{document}

\maketitle

%%%%%%%%%%%%%%%%%%%%%%%%%%%%%%%%%%%%%%%%%%%%%%%%%%%%%%%%%%%%%%%%%%%%%%%%%%%%%%%%
\begin{abstract}
A multi-source quickest detection problem is considered. Assume
there are two independent Poisson processes $X^{1}$ and $X^{2}$
with disorder times $\theta_{1}$ and $\theta_{2}$, respectively;
that is, the intensities of $X^1$ and $X^2$ change at random
unobservable times $\theta_1$ and $\theta_2$, respectively.
$\theta_1$ and $\theta_2$ are independent of each other and are
exponentially distributed. Define $\theta \triangleq \theta_1
\wedge \theta_2=\min\{\theta_{1},\theta_{2}\}$ . For any stopping
time $\tau$ that is measurable with respect to the filtration
generated by the observations define a penalty function of the
form
\[
R_{\tau}=\mathbb{P}(\tau<\theta)+c
\mathbb{E}\left[(\tau-\theta)^{+}\right],
\]
where $c>0$ and $(\tau-\theta)^{+}$ is the positive part of
$\tau-\theta$. It is of interest to find a stopping time $\tau$
that minimizes the above performance index. This performance
criterion can be useful for example in the following scenario:
There are two assembly lines that produce products $A$ and $B$,
respectively. Assume that the malfunctioning (disorder) of the
machines producing $A$ and $B$ are independent events. Later, the
products $A$ and $B$ are to be put together to obtain another
product $C$. A product manager who is worried about the quality of
$C$ will want to detect the minimum of the disorder times (as
accurately as possible) in the assembly lines producing $A$ and
$B$.  Another problem to which we can apply our framework is the
internet surveillance problem: A router receives data from, say,
$n$ channels. The channels are independent and the disorder times
of channels are $\theta_1, \cdots, \theta_n$. The router is said
to be under attack at $\theta=\theta_1 \wedge \cdots \wedge
\theta_n$. The administrator of the router is interested in
detecting $\theta$ as quickly as possible.

Since both observations $X^{1}$ and $X^{2}$ reveal information
about the disorder time $\theta$, even this simple problem is more
involved than solving the disorder problems for $X^{1}$ and
$X^{2}$ separately. This problem is formulated in terms of a three
dimensional sufficient statistic, and the corresponding optimal
stopping problem is examined. The solution is characterized by
iterating a suitable functional operator.
\end{abstract}



\pagestyle{myheadings} \thispagestyle{plain} \markboth{E.
BAYRAKTAR AND H. V. POOR}{Quickest Detection of a Minimum of Two
Poisson Disorder Times}

%%%%%%%%%%%%%%%%%%%%%%%%%%%%%%%%%%%%%%%%%%%%%%%%%%%%%%%%%%%%%%%%%%%%%%%%%%%%%%%%
\section{Introduction}
Consider two independent Poisson processes $X^{i}=\{X^{i}_t:t \geq
0\}$ $i \in \{1,2\}$ with the same arrival rate $\beta$. At some
random unobservable times $\theta_1$ and $\theta_2$, with
distributions
\begin{equation}\label{edist}
\P(\theta_i=0)=\pi_i \quad \P(\theta_i>t)=(1-\pi_i)e^{-\lambda
t}\,\, \text{for}\,\, t \geq 0,
\end{equation}
the arrival rates of the Poisson processes $X^1$ and $X^2$ change
from $\beta$ to $\alpha$, respectively, i.e.,
\begin{equation}\label{intensita}
X^i_{t}-\int_{0}^{t}h_i(s)ds, \quad t \geq 0,\,i=1,2,
\end{equation}
are martingales, in which
\begin{equation}
h_i(t)=[\beta 1_{\{s<\theta_i\}}+\alpha 1_{\{s \geq \theta_i\}}],
\quad t \geq 0, \, i=1,2.
\end{equation}
 Here $\alpha$ and $\beta$ are known
positive constants. We seek a stopping rule $\tau$ that detects
the instant $\theta=\theta_1 \wedge \theta_2$ of the first regime
change as accurately as possible given the past and the present
observations of the processes $X^1$ and $X^2$. More precisely, we
wish to choose a stopping time $\tau$ of the history of the
processes $X^1$ and $X^2$ that minimizes the following penalty
function
\begin{equation}\label{penaltyfunction}
R_{\tau}=\mathbb{P}(\tau<\theta)+c
\mathbb{E}\left[(\tau-\theta)^{+}\right].
\end{equation}
The first term in (\ref{penaltyfunction}) penalizes the frequency
of false alarms, and the second term penalizes the detection
delay. The disorder time demarcates two regimes, and in each of
these regimes the decision maker uses distinctly different
strategies. Therefore, it is in the decision maker's interest to
detect the disorder time as accurately as possible from its
observations. Here, we are solving the case when a decision maker
has two identical and independent sources to process. In the
Section~\ref{sec:extensions} we discuss how our analysis can be
extended to non-identical sources.

Quickest detection problems arise in a variety of applications
such as seismology, machine monitoring, finance, health, and
surveillance, among others (see e.g. \cite{nikiforov},
\cite{neftci}, \cite{mazumdar}, \cite{poor98} and
\cite{sonesson}). Because Poisson processes are often used to
model abrupt changes, Poisson disorder problems have potential
applications e.g. to the effective control and prevention of
infectious diseases, quickest detection of quality and reliability
problems in industrial processes, and surveillance of Internet
traffic to protect network servers from the attacks of malicious
users. This is because the number of patients infected, number of
defected items produced and number of packets arriving at a
network node are usually modeled by Poisson processes. In these
examples the disorder time corresponds to the time when an
outbreak occurs, when a machine in an assembly line breaks down or
when a router is under attack, respectively. The multi-source
quickest detection problem considered here  can be applied to
tackle these problems when there are multiple sources of
information. For example in the monitoring of industrial processes
the minimum of disorder times represents the first time when one
of many assembly lines in a plant breaks down during the
production of a certain type of item.  Let us be more specific:
Assume that there are two assembly lines that produce products $A$
and $B$, respectively. Assume also that the malfunctioning
(disorder) of the machines producing $A$ and $B$ are independent
events. Later, the products $A$ and $B$ are to be put together to
obtain another product $C$. A product manager who is worried about
the quality of $C$ will want to detect the minimum of the disorder
times (as accurately as possible) in the assembly lines producing
$A$ and $B$. The performance function (\ref{penaltyfunction}) is
an appropriate choice because the product manager will worry about
the quality of the end product $C$, not of the individual pieces
seperately. Another problem to which we can apply our framework is
the internet surveillance problem: A router receives data from,
say, $n$ channels. The channels are independent and the disorder
times of channels are $\theta_1, \cdots, \theta_n$. The router is
said to be under attack at $\theta=\theta_1 \wedge \cdots \wedge
\theta_n$. The administrator of the router is interested in
detecting $\theta$ as quickly as possible.

The one dimensional Poisson disorder problem, i.e., the problem of
detecting $\theta_1$ as accurately as possible given the
observations from the Poisson process $X^1$ has recently been
solved (see \cite{BD03}, \cite{BD04} and the references therein).
The two-dimensional disorder problem we have introduced cannot be
reduced to solving the corresponding one-dimensional disorder
problems since both $X^1$ and $X^2$ reveal some information about
$\theta$ whenever these processes jump. That is, if we take the
minimum of the optimal stopping times that solve the one
dimensional Poisson disorder problems, then we obtain a stopping
time that is a sub-optimal solution to (\ref{penaltyfunction})
(see Remark~\ref{stsubopti}).

We will show that the quickest detection problem of
(\ref{penaltyfunction}) can be reduced to an optimal stopping
problem for a three-dimensional piece-wise deterministic Markov
process. Continuous-time Markov optimal stopping problems are
typically solved by formulating them as free boundary problems
associated with the infinitesimal generator of the Markov process.
In this case, however the infinitesimal generator contains
differential delay operators. Solving free boundary problems
involving differential delay operators is a challenge even in the
one dimensional case and the smooth fit principle is expected to
fail (see \cite{BD03}, \cite{BD04} and the references therein).
Instead as in \cite{bdk05} and \cite{MR96b:90002} we work with an
integral operator, iteration of which generates a monotonically
increasing sequence of functions converging exponentially to the
value function of the optimal stopping problem. That is, using the
integral operator we reduce the problem to a sequence of
deterministic optimization problems. This approach provides a new
numerical method for calculating and characterizing the value
function and the continuation region in addition to providing
information about the shape and the location of the optimal
continuation region. Using the structure of the paths of the
piece-wise deterministic Markov process we also provide a
non-trivial bound on the optimal stopping time which can be used
to obtain approximate stopping strategies.

The remainder of this paper is organized as follows. In
Sections~\ref{probde} and ~\ref{optsttime}, we restate the problem
of interest under a suitable reference measure $\P_0$ that is
equivalent to $\P$. Working under the reference measure $\P_0$
reduces the computations considerably, since under this measure
the observations $X^1$ and $X^2$ are simple Poisson processes that
are independent of the disorder times. Here we show that the
quickest detection problem reduces to solving an optimal stopping
problem for a three-dimensional statistic.  In
Section~\ref{samppath}, we analyze the path behavior of this
sufficient statistic. In Section~\ref{reginbnd}, we provide a
tight upper bound on the continuation region of the optimal
stopping problem, which can be used to determine approximate
detection rules besides helping us to determine the location and
the shape of the continuation region. Here, we also show that the
smallest optimal stopping time of the problem under consideration
has finite expectation. In Section~\ref{vnsect}, we convert the
optimal stopping problem into sequences of deterministic optimal
stopping problems using a suitably defined integral operator. In
Section \ref{optstotimesect}, we construct optimal stopping times
from sequences of stopping (alarm) times that sound before the
processes $X^{1}$ and $X^2$ jump a certain number of times. In
Section \ref{stofopt} we discuss the structure of the optimal
stopping regions. And finally, we discuss how to extend our
approach to the case with more than two sources, and to the case
when the jump sizes are random and the jump size distribution
changes at the time of disorder.

\section{Problem Description}\label{probde}
Let us start with a probability space $(\Omega, \F, \P_0)$ that
hosts two independent Poisson processes $X^1$ and $X^2$, both of
which have rate $\beta$, as well as  two independent random
variables $\theta_1$ and $\theta_2$  independent of the Poisson
processes with distributions
\begin{equation}\label{eq:dist-pi}
\P_0(\theta_i=0)=\pi_i \quad \text{and} \quad
\P_0(\theta_i>t)=(1-\pi_i)e^{-\lambda_i t},
\end{equation}
for $0\leq t<\infty$, $i \in \{1,2\}$ and for some known constants
$\pi_i \in [0,1)$ and $\lambda>0$ for $i \in \{1,2\}$. We denote
by $\Fb=\{\F_t\}_{0 \leq t <\infty}$ the filtration generated by
$X^1$ and $X^2$, i.e., $\F_t=\sigma(X^1_s,X^2_s, 0 \leq s \leq
t)$, and denote by $\Gb=\{\G_t\}_{0 \leq t <\infty}$  the initial
enlargement of $\Fb$ by $\theta_1$ and $\theta_2$, i.e., $\G_t
\triangleq \sigma (\theta_1,\theta_2,X_s^{2},X_s^2: 0 \leq s \leq
t)$. The processes $X^1$ and $X^2$ satisfy (\ref{intensita}) under
a new probability measure $\P$, which is characterized by
\begin{equation}\label{defineP}
\frac{d\P}{d\P_0} \bigg |_{\G_t} \triangleq Z_t \triangleq Z^1_t
Z^2_t,
\end{equation}
where
\begin{equation}
Z^i_t \triangleq \exp\left(\int_0^t
\log\left(\frac{h_i(s-)}{\beta}\right)dX^i_s-\int_0^t
[h_i(s)-\beta]ds\right),
\end{equation}
for $t \geq 0$ and $i \in \{1,2\}$ are exponential martingales
(see e.g. \cite{MR82m:60058}). Under this new probability measure
$\mathbb{P}$ of (\ref{defineP}), $\theta_1$ and $\theta_2$ have
the same distribution as they have under the measure $\P_0$, i.e.,
their distribution is given by (\ref{edist}). This holds because
$\theta_1$ and $\theta_2$ are $\mathcal{G}_0$-measurable and
$d\P/d\P_0\big|_{\mathcal{G}_0}=1$, i.e., $\P$ and $\P_0$ coincide
on $\G_0$. Under the new probability measure $\mathbb{P}$ the
processes $X^1$ and $X^2$ have measurable intensities $h_1$ and
$h_2$ respectively. That is to say that (\ref{intensita}) holds.
In other words, the probability space $(\Omega, \mathcal{F},
\mathbb{P})$ describes the model posited in (\ref{edist}) and
(\ref{intensita}). Now, our problem is to find a quickest
detection rule for the disorder times $\theta_1 \wedge \theta_2$,
which is \emph{adapted to} the history $\mathcal{F}$ generated by
the observed processes $X^1$ and $X^2$ because the complete
information (concerning $\theta_1$ and $\theta_2$) embodied in
$\mathcal{G}$ is not available. We will achieve our goal by
finding an $\mathcal{F}$ stopping time that minimizes
(\ref{penaltyfunction}).

In terms of the exponential likelihood processes
\begin{equation}\label{likelihood}
L^i_{t} \triangleq \left(\frac{\alpha}{\beta}\right)^{X^i_t}
\exp(-(\alpha-\beta)t), \, t \geq 0,\, i \in \{1,2\},
\end{equation}
we can write
\begin{equation}\label{representationforZ}
Z_t^i=1_{\{\theta_i>t\}}+1_{\{\theta_i \leq t\}}
\frac{L^{i}_{t}}{L^{i}_{\theta_i}}
\end{equation}

Let us introduce the posterior probability process
\begin{equation}\label{posteriorprob}
\Pi_t \triangleq \P\left(\theta \leq t\big |
\F_t\right)=\frac{\E_0\left[Z_t 1_{\{\theta \leq t\}}\big| \F_t
\right]}{\E_0\left[Z_t\big| \F_t\right]},
\end{equation}
where the second equality follows from the Bayes formula (see e.g.
\cite{MR2001k:60001a}). Then it follows that from
(\ref{representationforZ}) and (\ref{posteriorprob}) that
\begin{equation}\label{1minuspit}
1-\Pi_t= \frac{(1-\pi)e^{-2\lambda t}}{\E_{0}\left[Z_t \big | \F_t
\right]}, \quad \text{where}
\end{equation}
\begin{equation}\label{eq:l-and-pi}
\pi \triangleq1-(1-\pi_1)(1-\pi_2).
\end{equation}
Let us now introduce  the odds-ratio process
\begin{equation}\label{eq:defn-phi-t}
\Phi_t \triangleq \frac{\Pi_t}{1-\Pi_t}, \,\, 0 \leq t <\infty.
\end{equation}
Then observe from (\ref{posteriorprob}) and (\ref{1minuspit}) that
\begin{equation}\label{relphiandz}
\E_0\left[Z_t 1_{\{\theta \leq
t\}}|\mathcal{F}_t\right]=(1-\pi)e^{-\lambda t} \Phi_t,
\end{equation}
$t \geq 0$. Now, we will write the penalty function of
(\ref{penaltyfunction}) in terms of the odds-ratio process.
\begin{equation}\label{cmp1}
\begin{split}
\E\left[(\tau-\theta)^{+}\right]&=\E\left[\int_{0}^{\infty}1_{\{\tau>t\}}1_{\{\theta
\leq t\}} dt\right]
=\int_{0}^{\infty}\E_0\left[1_{\{\tau>t\}}\E_0\left[Z_t
1_{\{\theta \leq t\}}\big| \F_t\right]\right]dt
\\&=(1-\pi)\E_0
\int_0^{\tau} e^{-2\lambda t} \Phi_t dt.
\end{split}
\end{equation}
Since $\{\tau < \theta\} \in \mathcal{G}_{\theta}$ we can write
\begin{equation}\label{cmp2}
\P(\tau<\theta)=\E_0\left[Z_{\theta}1_{\{\tau<\theta\}}\right]=\P_{0}(\tau<\theta)
=(1-\pi)\left(1-\lambda \E_{0}\left[\int_0^{\tau}e^{-2\lambda
t}dt\right]\right),
\end{equation}
where the second equality follows since $Z_{\theta}=1$ almost
surely under $P_{0}$. Using (\ref{cmp1}) and (\ref{cmp2}) we can
write the penalty function as
\begin{equation}\label{penalty}
R_{\tau}(\pi_1,\pi_2)=1-\pi+c(1-\pi)\E_0\left[\int_0^{\tau}e^{-2\lambda
t} \left(\Phi_t-\frac{\lambda}{c}\right)dt \right].
\end{equation}
On the other hand the following lemma obtains a representation for
the odds ratio process $\Phi$.
\begin{lemma}
Let us denote
\begin{equation}\label{eq:phi-t-i}
\Phi^{i}_t \triangleq \frac{e^{\lambda
t}}{1-\pi_i}\E_{0}\left[1_{\{\theta_i\leq
t\}}\frac{L^{i}_t}{L^{i}_{\theta_i}}\big|
\F^{i}_t\right]=\frac{\P(\theta_i \leq t|\F_t)}{1-\P(\theta_i \leq
t|\F_t)},
\end{equation}
for $t \geq 0$ and $i \in \{1,2\}$. Then we can write the
odds-ratio process $\Phi$  as
\begin{equation}
\Phi_t=\Phi^{1}_t+\Phi^{2}_t+\Phi^{1}_t \Phi^{2}_t,\quad t \geq 0
\end{equation}
\end{lemma}
\begin{proof}
From (\ref{relphiandz})
\begin{equation}
\begin{split}
 \Phi_t&=\frac{e^{2\lambda t}}{(1-\pi)}\E_0\left[Z_t 1_{\{\theta \leq
t\}}|\mathcal{F}_t\right],
\\&=\frac{e^{2\lambda t}}{(1-\pi)}\Bigg\{\P_0(\theta_1>t)\E_{0}\left[1_{\{\theta_1 \leq
t\}}\frac{L^{1}_t}{L^{1}_{\theta_1}}\big|
\F^{1}_t\right]+\P(\theta_2>t)\E_{0}\left[1_{\{\theta_2 \leq
t\}}\frac{L^{2}_t}{L^{2}_{\theta_i}}\big| \F^{i}_t\right]
\\&+\E_{0}\left[1_{\{\theta_1 \leq
t\}}\frac{L^{1}_t}{L^{1}_{\theta_1}}\big|
\F^{1}_t\right]\E_{0}\left[1_{\{\theta_2 \leq
t\}}\frac{L^{2}_t}{L^{2}_{\theta_2}}\big| \F^{2}_t\right] \Bigg\}
\end{split}
\end{equation}
The second equality follows from (\ref{defineP}),
(\ref{representationforZ}) and the independence of the sigma
algebras $\F_t^1$ and $\F_t^2$. Now the claim follows from
(\ref{eq:dist-pi}), (\ref{eq:l-and-pi}) and (\ref{eq:phi-t-i}).
\end{proof}

 Using the
fact that the likelihood ratio process $L^i$ is the unique
solution of the equation
\begin{equation}
dL^i_t=[(\alpha/\beta)-1]L^{i}_{t-}(dX^i-\alpha \,dt), \quad
L^i_0=1,
\end{equation}
(see e.g. \cite{RY99}) and by means of the chain-rule we obtain
\begin{equation}\label{eq:dynamics-phi-i}
d
\Phi^{i}_{t}=(\lambda+(\lambda-\alpha+\beta)\Phi^{i}_{t})dt+[(\alpha/\beta)-1]\Phi^{i}_{t}dX^{i}_t,
\quad \Phi^{i}_0=\frac{\pi_i}{1-\pi_i},
\end{equation}
for $t \geq 0$ and $i \in \{1,2\}$ (see \cite{BD04}). If we let
\begin{equation}\label{eq:defn-m-p-5}
\Phi^{+}_t \triangleq \Phi^{1}_{t}+\Phi^{2}_t,\,\,\,
\Phi^{\times}_t \triangleq \Phi^{1}_{t} \Phi^{2}_t, \,\,\,t \geq
0,
\end{equation}
then using a change of variable formula for jump processes gives
\begin{equation}\label{dynpm}
\begin{split}
d \m_t&= [\lambda \p_t+2a \m_t]dt+ (\alpha/\beta)-1)\m_t
d(X^1_t+X^2_t),
\\ d\Phi^{+}_{t}&=[2 \lambda+
a\Phi^{+}_t]dt+((\alpha/\beta)-1)[\Phi^{1}_t dX^1_t+\Phi^{2}_t
dX^2_t]
\end{split}
\end{equation}
with $\m_0=\pi_1 \pi_2/[(1-\pi_1)(1-\pi_2)]$, and
$\p_0=\pi_1/(1-\pi_1)+\pi_2/(1-\pi_2)$, where $a \triangleq
\lambda-\alpha+\beta$. Note that $X_{t} \triangleq
X^{1}_t+X^{2}_t$, $t \geq 0$, is a Poisson process with rate $2
\beta$ under $\P_0$.

It is clear from (\ref{eq:dynamics-phi-i}) and (\ref{dynpm}) that
\begin{equation}\label{eq:suff-stat}
\Upsilon \triangleq (\m,\p,\Phi^{1}),
\end{equation}
 is a
piece-wise deterministic Markov process; therefore the original
change detection problem with penalty function
(\ref{penaltyfunction}) has been reformulated as (\ref{penalty})
and (\ref{eq:dynamics-phi-i})-(\ref{eq:suff-stat}), which is an
optimal stopping problem for a two dimensional Markov process
driven by three dimensional piecewise-deterministic Markov
process.

We will denote by $\mathcal{A}$ the infinitesimal generator of
$\Upsilon$. Its action on a smooth test function $f:\mathbb{B}^3_+
\rightarrow \mathbb{R}$ is given by
\begin{equation}\label{eq:infinitesimall}
\begin{split}
&[\mathcal{A}f](\ms,\ps,\phi^1)=D_{\ms}f(\ms,\ps,\phi^1)[\lambda
\ps+2 a \ps]+D_{\ps}f(\ms,\ps,\phi^1)[2 \lambda + a
\ps]\\&+D_{\phi^1}f(\ms,\ps,\phi^1)[\lambda+a
\phi^1]+\beta\left[f\left(\frac{\alpha}{\beta}\, \ms,
\ps+\left(\frac{\alpha}{\beta}-1\right)\phi^1,\frac{\alpha}{\beta}\,\phi^1
\right)-f(\ms,\ps,\phi^1)\right]
\\&+\beta\left[f\left(\frac{\alpha}{\beta}\, \ms,
\frac{\alpha}{\beta}\,\ps-\left(\frac{\alpha}{\beta}-1\right)\phi^1,\phi^1
\right)-f(\ms,\ps,\phi^1)\right].
\end{split}
\end{equation}

Let us denote
\begin{equation} \mathbb{B}^{2}_+ \triangleq \{(x,y)
\in \mathbb{R}^{2}_+: y \geq 2 \sqrt{x} \} \quad \text{and}
\end{equation}
\begin{equation} \mathbb{B}^3_+ \triangleq \{(x,y,z)
\in \mathbb{R}^{3}_+: y \geq 2 \sqrt{x},y \geq z \}.
\end{equation}

  Now, for every $(\ms, \ps, \phi^1) \in \mathbb{B}_{+}^3$, let
us denote denote by $x(t,\ms)$, $y(t,\ps)$ and $z(t,\phi^1)$, $t
\in \mathbb{R}$, the solutions of
\begin{equation}\label{diffeqn}
\begin{split}
\frac{d}{dt}x(t,\ms)&=[\lambda y(t,\ps)+2 a x(t,\ms)]dt, \quad
x(0,\ms)=\ms,
\\ \frac{d}{dt}y(t,\ps)&=[2 \lambda+ a y(t,\ps)]dt, \quad
y(0,\ps)=\ps,
\\ \frac{d}{dt}z(t,\phi^1)&=[\lambda+a z(t,\phi^1)]dt, \quad
z(0,\phi^1)=\phi^1.
\end{split}
\end{equation}
The solutions of (\ref{diffeqn}), when $a \neq 0$, are explicitly
given by
\begin{equation}\label{detversion}
\begin{split}
x(t,\ms)&=\frac{ \lambda^2}{a^2}+e^{2at}\left[\ms-\frac{ \lambda^2
}{a^2}\right]+e^{2 a t} (1-e^{-at})
\frac{\lambda}{a}\left(\ps+\frac{2 \lambda}{a}\right),
\\ y(t,\ps)&=-\frac{2 \lambda}{a}+e^{at}\left(\ps+\frac{2
\lambda}{a}\right),
\\ z(t,\phi^1)&=-\frac{ \lambda}{a}+e^{at}\left(\ps+\frac{
\lambda}{a}\right).
\end{split}
\end{equation}
Otherwise, $x(t,\ms)=\ms+\lambda t \ms+\lambda^2 t^2$,
$y(t,\ps)=\ps+2\lambda t$, and $z(t,\phi^1)=\phi^1+\lambda t$.
 Note that the solution $(x,y,z)$ of the
system of equations in (\ref{diffeqn}) satisfy the semi-group
property, i.e., for every $s,t \in \mathbb{R}$
\begin{equation}\label{semigroup}
x(t+s,\phi_0)=x(s,x(t,\phi_0)), \quad
y(t+s,\phi_1)=y(s,y(t,\phi_1)), \quad \text{and} \quad
z(t+s,\phi_1)=z(s,z(t,\phi_1)).
\end{equation}

Note from (\ref{eq:dynamics-phi-i}), (\ref{dynpm}) and
(\ref{detversion}) that
\begin{equation}
\m_t=x(t-\sigma_n,\m_{\sigma_n}), \,\,
\p_t=y(t-\sigma_n,\p_{\sigma_n}),\,\,
\Phi^1_t=z(t-\sigma_n,\Phi^1_{\sigma_n}) \quad \sigma_n \leq
t<\sigma_{n+1}, \, n\in \mathbb{N},
\end{equation}
and {\small
\begin{equation}
\begin{split}
 \m_{\sigma_{n+1}}&=\frac{\alpha}{\beta}\,\m_{\sigma_{n+1-}},
 \quad \Phi^1_{\sigma_{n+1}}=\frac{\alpha}{\beta}\,\Phi^1_{\sigma_{n+1}}1_{\{X^{1}_{\sigma_{n+1}}\neq X^{1}_{\sigma_{n+1-}
}\}}, \quad \text{and},
\\ \p_{\sigma_{n+1}}&=\left[\p_{\sigma_{n+1}-}+\left(\frac{\alpha}{\beta}-1\right)\Phi^1_{\sigma_{n+1-}}\right]
1_{\{X^{1}_{\sigma_{n+1}}\neq X^{1}_{\sigma_{n+1-}
}\}}+\left[\frac{\alpha}{\beta} \,
\p_{\sigma_{n+1}-}-\left(\frac{\alpha}{\beta}-1\right)\Phi^1_{\sigma_{n+1-}}\right]
1_{\{X^{2}_{\sigma_{n+1}}\neq X^{2}_{\sigma_{n+1-} }\}}
\end{split}
\end{equation}
} Here, for any function $h$, $h(t-) \triangleq \lim_{s \uparrow
t}h(t)$. Note that an observer watching $\Upsilon$ is able to tell
whenever the processes $X^1$ and $X^2$ jump (see
(\ref{eq:dynamics-phi-i}), (\ref{dynpm})), i.e., the filtration
generated by $\Upsilon$ is the same as $\mathbb{F}$.

\section{An Optimal Stopping Problem}\label{optsttime}

Let us denote  the set of $\Fb$ stopping times by $\mathcal{S}$.
  The value function of the quickest detection problem
\begin{equation}
U(\pi_1,\pi_2) \triangleq \inf_{\tau \in \mathcal{S}}
R_{\tau}(\pi_1,\pi_2)
\end{equation}
can be written as
\begin{equation}
U(\pi_1,\pi_2)=(1-\pi) \left[1+cV\left(\frac{\pi_1 \pi_2}{1-\pi}
,\frac{\pi_1+\pi_2-2 \pi_1
\pi_2}{1-\pi},\frac{\pi_1}{1-\pi_1}\right)\right],
\end{equation}
where $V$ is the value function of the optimal stopping problem
\begin{equation}\label{optstoppingprob}
V(\ms,\ps,\phi^1) \triangleq \inf_{\tau \in
\mathcal{S}}\E^{(\ms,\,\ps,\,\phi^1)}_0\left[\int_{0}^{\tau}e^{-\lambda
t} h\left(\m_t,\p_t\right)dt\right],
\end{equation}
in which $(\ms,\ps,\phi^1)\in \mathbb{D}^{3}_{+}$, and $h(x,y)
\triangleq x+y-\lambda/c$.  Here $\E^{(\ms,\,\ps,\,\phi^1)}_0$ is
the expectation under $\P_0$ given that $\m_0=\ms$, $\p_0=\ps$,
$\Phi^{1}_0=\phi^1$.

It is clear from (\ref{optstoppingprob}) that for both optimal
stopping problems it is not optimal to stop before $(\m_t,\p_t)$,
$t \geq 0$, leaves the \emph{advantageous region} defined by
\begin{equation}\label{co}
\mathbb{C}_0 \triangleq \{(\ms,\ps) \in \mathbb{B}^{2}_+: \ms+\ps
\leq \lambda/c\}.
\end{equation}
Let us also denote
\begin{equation}\label{co-tilde}
\mathbb{C} \triangleq \{(\ms,\ps,\phi^1) \in \mathbb{B}^3_+:
\ms+\ps \leq \lambda/c\}.
\end{equation}

 Also note that the only reason not to stop at the time of
the first exit from the region $\mathbb{C}_{0}$ is the prospect of
$(\m_t,\p_t)$, $t \geq 0$ returning to $\mathbb{C}_0$ at a future
time.

\begin{remark}
It is reasonable to question our choice of statisitc, since it is
clear that $(\Phi^1_t,\Phi^2_t)_{t \geq 0}$ contains all the
information $X$ has to offer. Our choice $(\Upsilon_t)_{t \geq
0}$, which is defined in (\ref{eq:suff-stat}), is motivated by the
mere desire of having a concave value function $U$ and a convex
optimal stopping region. The concavity is due to the linearity of
the function $h$ (see Lemma~\ref{proofJ} and its proof along with
Lemmas~\ref{expofast} and (\ref{defnofvn}), and
Theorem~\ref{vneVn}).

If we had chosen to work with $(\Phi^1_t,\Phi^2_t)_{t \geq 0}$,
then the relevant optimal stopping problem becomes
\begin{equation}\label{optstoppingprob-t}
W(a,b) \triangleq \inf_{\tau \in
\mathcal{S}}\E^{(a,b)}_0\left[\int_{0}^{\tau}e^{-\lambda t}
\tilde{h}\left(\Phi^1_t,\Phi^2_t\right)dt\right],
\end{equation}
in which
\begin{equation}
\tilde{h}(x,y) \triangleq x+y+xy, \quad (x,y) \in \mathbb{R}_+^2.
\end{equation}
Since $h(\cdot, \cdot)$ is non-linear the concavity of the value
function, i.e. $W(\cdot,\cdot)$ is not concave. In fact,
$W(x,y)=V(xy,x+y,x)$. The function $V$ is concave but $W$ is not.
So there is a trade off between concavity and the dimension of the
statistic to be used.
\end{remark}

\emph{In what follows, for the sake of the simplicity of notation,
when the meaning is clear, we will drop the superscripts of the
expectation operators $\E^{(\ms,\,\ps,\,\phi^1)}_0$.}

\section{Sample Paths of $\Psi=(\m,\p)$}\label{samppath}
It is illustrative to look at the sample paths of the sufficient
statistic $\Psi$, to understand the nature of the problem. Indeed,
this way, for a certain parameter range, we will be able to
identify the optimal stopping time without any further analysis.

From (\ref{detversion}), we see that, if $a>0$, then the paths of
the processes $\m$ and $\p$ increase between the jumps, and
otherwise the paths of the processes $\m$ and $\p$ mean-revert to
the levels $2 \lambda^2/a^{2}$ and $-2\lambda/a$ respectively.
Also observe that $\m$, $\p$ increase (decrease) with a jump if
$\alpha \geq \beta$ ($\beta>\alpha$). See (\ref{dynpm}).

\emph{Case I: $\alpha \geq \beta$, $a>0$.} The following theorem
follows from the description of the behavior of the paths above.
\begin{thm}\label{thm:first}
If $\alpha>\beta$ and $a>0$, then the stopping rule
\begin{equation}
\tau_0 \triangleq \inf\{t \geq 0: \m_t+\p_t \geq \lambda/c \}
\end{equation}
is optimal for (\ref{optstoppingprob}).
\end{thm}
\begin{proof}
Under the hypothesis of the theorem and whenever a path of
$(\m,\p)$ leaves $\mathbb{C}_0$ it never returns.
\end{proof}

 In Section~\ref{reginbnd}, we will identify another case (another range of parameters)
 in which the advantageous region
$\mathbb{C}_{0}$ is the optimal continuation region and the
stopping time $\tau_{0}$ is optimal (see Cases II-b-i-1 and
II-b-ii-1).
\begin{remark}\label{stsubopti} Let $\kappa_i
\triangleq \inf \{t \geq 0:\Phi_{t}^{i} \geq \lambda/c\}$. If
$\alpha \geq \beta$ and $a>0$, then $\kappa_i$ is the optimal
stopping time for the one dimensional disorder problem with
disorder time $\theta_i$ (\cite{BD04}). Let us define $\kappa
\triangleq \kappa_1 \wedge \kappa_2$. Since with this choice of
parameters $\m_{\kappa}+\p_{\kappa}>\lambda/c$, it follows that
$\tau_0< \kappa$ almost surely. Therefore, if we follow the rule
dictated by the stopping time $\kappa$, then we pay an extra
penalty for detection delay. This example illustrates that solving
the two one dimensional quickest detection problems separately in
order to minimize the penalty function of (\ref{penaltyfunction})
is suboptimal.
\end{remark}

In what follows, we will consider the remaining cases: $\alpha
\geq  \beta$ and $a<0$; $\alpha<\beta$.

\section{Construction of a Bound on the Continuation
Region}\label{reginbnd}

In this section the purpose is to show that the continuation
region of (\ref{optstoppingprob}) is bounded. The construction of
upper bounds is carried out in the next two theorems. These upper
bounds are tight as the next theorem shows and might be used to
construct useful approximations to the two optimal stopping times
solving the problems defined in (\ref{optstoppingprob}). We will
carry out the analysis for $a=\lambda-\alpha+\beta \neq 0$. A
similar analysis for this case can be similarly performed. As a
result of Theorem~\ref{thm:finite-expectation} we are also able to
conclude that the (smallest) optimal stopping time has a finite
expectation.

The first two theorems in this section assume that an optimal
stopping time of (\ref{optstoppingprob}) exists and in particular
the stopping time
\begin{equation}\label{optst}
\tau^{*}(a,b,c) \triangleq \inf\{t \geq 0: V(\Upsilon_t)=0,\,
\Upsilon_0=(a,b,c)\},
\end{equation}
is optimal.  In Section~\ref{optstotimesect}, we will verify that
this assumption holds. We will denote by
\begin{equation}\label{eq:srop-regn}
\mathbf{\Gamma} \triangleq \{(a,b,c) \in \mathbb{B}^{3}_{+}:
v(a,b,c)=0\}, \quad \mathbf{C} \triangleq
\mathbb{B}^{3}_{+}-\mathbf{\Gamma},
\end{equation}
the optimal stopping region and optimal continuation region of
(\ref{optstoppingprob}) respectively.

\begin{thm}\label{thm:sec-5-ilk}
In this theorem the standing assumption is that $\alpha \geq
\beta$ and that $a<0$ (Case II).

(Case II-a) Let us further assume that $\lambda/a^2-2/a \leq 1/c$
and denote
\begin{equation}
\mathbb{D}_0 \triangleq \left\{(x,y) \in \mathbb{B}^{2}_+:x \cdot
\left(\frac{1}{2 (\lambda-a)}\right) +y \cdot\left(\frac{3
\lambda-2 a }{2 (\lambda-a)(2 \lambda-a)}\right)+k
>0\right\},
\end{equation}
in which
\begin{equation}\label{eq:k}
k \triangleq
\frac{\lambda}{2a^2}-\frac{1}{a}-\frac{1}{2c}+\frac{\lambda^2}{2
a^2 (\lambda-a)}+\frac{1}{2\lambda-a}+2\left(\frac{\lambda}{a}-
\frac{\lambda^2}{a^2} \right).
\end{equation}
Let $(\phi_0,\phi_1) \in \mathbb{D}_0 \cap
(\mathbb{B}^2_+-\mathbb{C}_0)$. Then for any $\phi_2 \leq \phi_1$,
$(\phi_0,\phi_1,\phi_2)$ is in the stopping region of
(\ref{optstoppingprob}).

 (Case II-b) Assume that $\lambda/a^2-2/a \geq 1/c$ (standing assumption in the rest of the theorem).
Consider the four different possible ranges of parameters:

Case II-b-i: If $\lambda+a \leq 0$
\begin{itemize}
\item and if $-a/c-1 \leq 0$ (Case II-b-i-1), then
$\mathbb{C}$ in (\ref{co-tilde}) is the optimal continuation
region for (\ref{optstoppingprob}).

\item  Else if $-a/c-1>0$ (Case II-b-i-2), then a superset of the continuation region can be constructed as follows.
 Let us introduce the line segment
\begin{equation}\label{C}
C \triangleq \{(x,y) \in \mathbb{B}^{2}_+: x+y=\lambda/c\},
\end{equation}
and define
\begin{equation}
\begin{split}
C_{1}  \triangleq & \left\{(x,y) \in \mathbb{B}^{2}_{+}:
x=x(t,x^*), y=y(t,0) \,\, \text{for} \,\, t \in [0,t^*]  \right\}
\\ & \bigcup
\left\{(x,y) \in C: x<\frac{\lambda}{c}+\frac{2 \lambda
}{\lambda-a}\left(1+\frac{a}{c}\right),\,y>-\frac{2 \lambda
}{\lambda-a}\left(1+\frac{a}{c}\right)\right\}.
\end{split}
\end{equation}
Here, $t^{*}$ is the solution of $y(-t^*,-\frac{2 \lambda
}{\lambda-a}\left(1+\frac{a}{c}\right))=0$ and
$x^*=x(-t^*,\frac{\lambda}{c}+\frac{2 \lambda
}{\lambda-a}\left(1+\frac{a}{c}\right))$ The curve $C_{1}$
separates $\mathbb{R}^{2}_{+}$ into two connected regions. Let us
denote the region that lies above the curve $C_1$ by
\begin{equation}\label{regd1}
\mathbb{D}_{1} \triangleq \{(x,y) \in \mathbb{B}^{2}_+:\,
\text{there exists a positive number $\tilde{y}(x)<y$ such that}\,
(x,\tilde{y}(x)) \in C_1\}.
\end{equation}
Then  $ [(\mathbb{B}^{2}_{+}-\mathbb{D}_1) \times \mathbb{R}_+]
\cap \mathbb{B}^3_+$ is an upper bound on the continuation region
of  (\ref{optstoppingprob}).
\end{itemize}

Case II-b-ii: If $\lambda+a>0$
\begin{itemize}
\item and if $-a/c-1<0$ (Case II-b-ii-1), then $\mathbb{C}$ in (\ref{co-tilde}) is the optimal continuation
region for (\ref{optstoppingprob}).
\item Else if $-a/c-1>0$, then  $ [(\mathbb{B}^{2}_{+}-\mathbb{D}_1)
\times \mathbb{R}_+] \cap \mathbb{B}^3_+$ is an upper bound on the
continuation region  of  (\ref{optstoppingprob}).
\end{itemize}
Note that all the supersets of the continuation we constructed are
bounded subsets of $\mathbb{R}^3_+$.
\end{thm}

\begin{proof}

Note that
\begin{equation}\label{pgu}
\p_t \geq y(t,\phi_1), \quad \m_t \geq x(t,\phi_0), \,\, t \geq 0,
\end{equation}
almost surely if $\p_0=\phi_1$ and $\m_0=\phi_0$. This is because
$\m$ and $\p$ increase with jumps.

From this observation we obtain the following inequality
\begin{align}
&\inf_{\tau \in
\mathcal{S}}\E_{0}\left[\int_{0}^{\tau}e^{-2\lambda
s}h(\m_s,\p_s)ds\right] \label{ndetopts} \geq \inf_{\tau \in
\mathcal{S}}\E_{0}\left[\int_{0}^{\tau}e^{-2\lambda
s}h(x(s,\phi_0),y(s,\phi_1))ds\right]
\\&=\inf_{t \in [0,\infty]} \int_{0}^{t} e^{-2\lambda
s}h(x(s,\phi_0),y(s,\phi_1))ds. \label{detoptstop}
\end{align}

Note that if for a given $(\phi_0,\phi_1)$ the expression in
(\ref{detoptstop}) is equal to zero, then the infimum on the left
hand side of (\ref{ndetopts}) is attained by setting $\tau=0$. In
what follows we will find a subset of the stopping region of the
optimal stopping problem using this argument.

\emph{Case II-a: $\lambda/a^2-2/a \leq 1/c$.} In this case the
mean reversion level of the path $(x(\cdot,\phi_0),
y(\cdot,\phi_1))$, $(\phi_0,\phi_1)\in \mathbb{B}^2_+$, namely
$(\lambda^2/a^2,-2\lambda/a)$, is inside the region $\mathbb{C}_0$
which is defined in (\ref{co}). In this case, for any
$(\phi_0,\phi_1) \in \mathbb{B}^2_+-\mathbb{C}_0$ the minimizer
$t_{\text{opt}}(\phi_0,\phi_1)$ of the expression in
(\ref{detoptstop}) is either 0 or $\infty$ by the following
argument. For any $(\phi_0,\phi_1) \in
\mathbb{B}^{2}_{+}-\mathbb{C}_0$ the path
$(x(\cdot,\phi_0),y(\cdot,\phi_1))$ is in the advantageous region
$\mathbb{C}_{0}$ all the time except for possibly a finite
duration. Therefore if
\begin{equation}\label{otoinfty}
\int_{0}^{\infty}  e^{-2\lambda s}h(x(s,\phi_0),y(s,\phi_1))ds<0,
\end{equation}
then in order to minimize (\ref{detoptstop}) it is never optimal
to stop. On the other hand if (\ref{otoinfty}) is positive, then
it is not worth taking the journey into the advantageous region
and it is optimal to stop immediately in order to minimize
(\ref{detoptstop}).

We shall find the pairs $(\phi_0,\phi_1)$ for which
$t_{\text{opt}}= 0$. Using (\ref{detversion}) we can write
\begin{equation}
\int_{0}^{\infty}  e^{-2\lambda s}h(x(s,\phi_0),y(s,\phi_1))ds
=\phi_0 \left(-\frac{1}{\alpha-\beta}\right) + \phi_1
\left(\frac{a}{(\alpha-\beta)^2}\right)+k,
\end{equation}
where $k$ is given by (\ref{eq:k}).  Note that if $(\phi_0,\phi_1)
\in  \mathbb{D}_0 \cap (\mathbb{B}^2_+-\mathbb{C}_0)$, then by
(\ref{ndetopts}) and (\ref{detoptstop}) we can see that the
infimum in (\ref{detoptstop}) is equal to 0. Therefore
$[((\mathbb{B}^2_{+}-\mathbb{D}_0) \cup \mathbb{C}_0)\times
\mathbb{R}_+]\cap \mathbb{B}_+^3$  is a superset of the optimal
continuation region of (\ref{optstoppingprob}).

\emph{Case II-b: $\lambda/a^2-2/a \geq 1/c$.} In this case the
mean reversion level of $t \rightarrow (x(t,\phi_0), y(t,\phi_1))$
is outside $\mathbb{C}_0$. Therefore, the minimizer of
(\ref{detoptstop}) is $t_{\text{opt}}(\phi_0,\phi_1) \in \{0,
t_{c}(\phi_0,\phi_1), \infty \}$ where $t_c(\phi_0,\phi_1)$ is the
exit time of the path $(x(t,\phi_0),y(t,\phi_1))$ from
$\mathbb{C}_{0}$. The derivative
\begin{equation}\label{derivative}
\frac{d}{dt}[x(t,\phi_0)+y(t,\phi_1)]=(\lambda+a)y(t,\phi_1)+2a
x(t,\phi_1)+2 \lambda
\end{equation}
 vanishes if
$(x(t,\phi_0),y(t,\phi_1))$ meets the line segment
\begin{equation}
L=\{(x,y) \in \mathbb{B}^{2}_+: (\lambda+a)y+2ax+ 2 \lambda=0\}.
\end{equation}
Note that the mean reversion level belongs to $L$, i.e.,
\begin{equation}\label{meanreversion}
\left(\frac{\lambda^2}{a^2},-\frac{2\lambda}{a}\right)\in L.
\end{equation}

\emph{Case II-b-i: $\lambda+a<0$.} (In addition to $\alpha>\beta$,
$a<0$ and $\lambda/a^2-2/a \geq 1/c$.) In this case the line $L$
is decreasing (as a function of $x$).

\emph{Case II-b-i-1: $-a/c-2<0$.} (In addition to $\alpha>\beta$,
$a<0$, $\lambda/a^2-2/a \geq 1/c$ and $\lambda+a<0$.) In this case
the line segment $C$ in (\ref{C}) lies entirely below $L$. Assume
that a path $(x(\cdot,\phi_0),y(\cdot,\phi_1))$ originating at
$(\phi_0,\phi_1) \in \mathbb{B}^{2}_+-\mathbb{C}_0$ enters
$\mathbb{C}_0$ at time $t_0>0$. This path must leave
$\mathbb{C}_0$ at time $t_1<\infty$ since the mean reversion level
$( \lambda^2 / a^2, -\lambda/a ) \notin \mathbb{C}_0$. This
implies that for any $t \in (t_0,t_1)$
$x(t,\phi_0)+y(t,\phi_1)<\lambda/c$ and
$x(t_0,\phi_0)+y(t_0,\phi_1)=\lambda/c$. This yields a
contradiction, because $\lambda+a<0$ together with
(\ref{derivative}) implies that $t \rightarrow
x(t,\phi_0)+y(t,\phi_1)$ is increasing below the line segment $L$.
Therefore the minimizer $t_{\text{opt}}(\phi_0,\phi_1)$ of
(\ref{detoptstop}) is equal to 0 if $(\phi_0,\phi_1) \notin
\mathbb{C}_0$, and it is equal to $t_{c}(\phi_0,\phi_1)$ if
$(\phi_0,\phi_1) \in \mathbb{C}_0$. From (\ref{ndetopts}) we can
conclude that $\mathbb{C}$ is equal to the optimal continuation
region of (\ref{optstoppingprob}).

\emph{Case II-b-i-2}: $-a/c-1>0$. In this case the line segments
$C$ and $L$ intersect at
$I=(x^{I},y^{I})\triangleq(\frac{\lambda}{c}+\frac{2 \lambda
}{\lambda-a}\left(1+\frac{a}{c}\right),-\frac{2 \lambda
}{\lambda-a}\left(1+\frac{a}{c}\right))$. By running the paths
backward in time, we can find $x^*$ such that
\begin{equation}
(x^*,0)=\left(x\left(-t^*,x^{I}\right), y\left(-t^*,
y^{I}\right)\right).
\end{equation}
By the semi-group property (\ref{semigroup}), we have
\begin{equation}
\begin{split}
x(t^*,x^*)&=x\left(t^*,x\left(-t^*,x^I\right)\right)=x\left(t^*+(-t^*),x^I\right)
\\ &= x\left(0, x^I\right)=x^I.
\end{split}
\end{equation}
Similarly, $y(t^*,0)=y^I$. The function $t \rightarrow
x(t,x^*)+y(t,0)$ is decreasing on $(0,t^*)$ and increasing on
$(t^*,\infty)$. It follows that the path $t \rightarrow
(x(t,x^*),y(t,0))$ is tangential to $C$ at $I$ and lies above the
region $\mathbb{C}_0$.

 We will now show that if a path
$(x(\cdot,\phi_0), y(\cdot,\phi_1))$ originates in $\mathbb{D}_1$,
then it stays in $\mathbb{D}_1$. Let us first consider a pair
$(\phi_0,\phi_1) \in \mathbb{D}_{1}$ such that
$\phi_1<-2\lambda/a$. Consider the curve
\begin{equation}
P \triangleq \left \{(x,y) \in \mathbb{B}^{2}_+: x=x(t,x^*),
y=y(t,0) \,\, \text{for} \,\, t \in [0, \infty) \right\}.
\end{equation}
The following remark will be useful in completing the proof.
\begin{remark}\label{nointersection}
The semi-group property in (\ref{semigroup}) implies that two
distinct curves $(x(\cdot,\phi^a_0)$ $,y(\cdot,\phi^a_1))$ and
$(x(\cdot,\phi^b_0),y(\cdot,\phi^b_1))$ do not intersect. If
\begin{equation}
(x(t^a,\phi^a_0),y(t^a,\phi^a_1))=(x(t^b,\phi^b_0),y(t^b,\phi^b_1))=(\phi_0,\phi_1)
\end{equation}
for some $t^a, t^b \in \mathbb{R}$ then (\ref{semigroup}) implies
that
\begin{equation}
\begin{split}
(x(t,\phi^a_0),y(t,\phi^a_1))&=(x(t^a+(t-t^a),\phi^a_0),y(t^a+(t-t^a),\phi^a_1))
\\&=(x(t-t^a,\phi_0),y(t-t^a,\phi_1))=(x(t^b+(t-t^a),\phi^b_0),y(t^b+(t-t^a),\phi^b_1))
\\&=(x(t^b-t^a+t,\phi^b_0),y(t^b-t^a+t,\phi^b_1)), \quad \text{for
all} \quad t \in \mathbb{R},
\end{split}
\end{equation}
i.e., the two curves are identical after a reparametrization.
\end{remark}

If the point $(\phi_0,\phi_1)$ lies above $P$, and we recall that
$P$ lies above $\mathbb{C}_0$, then by Remark~\ref{nointersection}
the path $(x(\cdot,\phi_0), y(\cdot,\phi_1))$ will lie above
$\mathbb{C}_0$. If the point $(\phi_0,\phi_1)$ lies between $P$
and $\mathbb{C}_0$, then the path $(x(\cdot,\phi_0),
y(\cdot,\phi_1))$ will lie below the line segment $L$. This
observation together with the fact that $\lambda+a<0$,
 implies (using (\ref{derivative})) that the function $t \rightarrow
x(t,\phi_0)+y(t,\phi_1)$ is increasing. Therefore the path
$(x(\cdot,\phi_0), y(\cdot,\phi_1))$ cannot intersect
$\mathbb{C}_0$.

Now let us consider a pair $(\phi_0,\phi_1) \in \mathbb{D}_{1}$
such that $\phi_1>-2\lambda/a$. If $(\phi_0,\phi_1)$ lies above
$L$, then the function $t \rightarrow x(t,\phi_0)+y(t,\phi_1)$ is
decreasing and its range is $[\phi_0+\phi_1, 2 \lambda/a
(\lambda/a-1))$, which is always above $\lambda/c$, and therefore
the path $(x(\cdot,\phi_0),y(\cdot,\phi_1))$ does not enter
$\mathbb{C}_0$. If the point $(\phi_0,\phi_1)$ lies below $L$ then
$t \rightarrow x(t,\phi_0)+y(t,\phi_1)$ is increasing. This
monotonicity implies that the path
$(x(\cdot,\phi_0),y(\cdot,\phi_1))$ cannot visit $\mathbb{C}_0$.
If $\phi_1=-2\lambda/a$, then $y(t,\phi_1)=-2 \lambda/a$ for all
$t \geq 0$. $x(t,\phi_0)$ increases or decreases depending on
whether $(\phi_0,-2\lambda/a)$ is below or above $L$. Therefore if
$(\phi_0,-2\lambda/a) \notin \mathbb{C}_0$, then
$(x(\cdot,\phi_0),y(\cdot,\phi_1))$ never visits $\mathbb{C}_0$.
These arguments show that if a path $(x(\cdot,\phi_0),
y(\cdot,\phi_1))$ originates in $\mathbb{D}_1$, then it stays in
$\mathbb{D}_1$. Therefore if $(\phi_0,\phi_1) \in \mathbb{D}_1$,
then the infimum in (\ref{detoptstop}) is equal to 0 (by
(\ref{ndetopts}) and (\ref{detoptstop})). Therefore
$[(\mathbb{B}^{2}_+-\mathbb{D}_1)\times \mathbb{R}_+] \cap
\mathbb{B}^3_+$ is a superset of the optimal continuation region
of (\ref{optstoppingprob}).

\emph{Case II-b-ii: $\lambda+a>0$.} In this case $L$ is increasing
(as a function of $x$). The function $t \rightarrow
x(t,\phi_0)+y(t,\phi_1)$ is increasing if $(\phi_0,\phi_1)$ lies
above $L$, and it is decreasing otherwise.

\emph{Case II-b-ii-1: $-a/c-1 \leq 0$.} In this case the line
segments $L$ and $C$ do not intersect.
 Let us first consider a pair $(\phi_0,\phi_1) \in
\mathbb{B}^{2}_+$ such that $\phi_1<-2\lambda/a$. If
$(\phi_0,\phi_1) \notin \mathbb{C}_0$ lies above the line segment
$L$, then  $t \rightarrow x(t,\phi_0)+y(t,\phi_1)$ is increasing
and the path $(x(\cdot,\phi_0),y(\cdot,\phi_1))$ cannot enter
$\mathbb{C}_0$. Consider the curve
\begin{equation}
\tilde{P} \triangleq \left \{(x,y) \in \mathbb{B}^{2}_+:
x=x\left(t,-\frac{\lambda}{a}\right), y=y(t,0) \,\, \text{for}
\,\, t \in [0, \infty) \right\},
\end{equation}
which starts at the intersection of $L$ with the $x$-axis. The
semi-group property Remark~\ref{nointersection} implies that no
path starting to the right of $\tilde{P}$ intersects $\tilde{P}$
and therefore lies to the right of the region $\mathbb{C}_0$.
Therefore, if $(\phi_0,\phi_1)$ is below the line segment $L$,
then the path $(x(\cdot,\phi_0),y(\cdot,\phi_1))$ never visits the
advantageous region $\mathbb{C}_0$. (Note that if the path
$(x(\cdot,\phi_0),y(\cdot,\phi_1))$ meets the line $L$ at time
$t_{L}(\phi_0,\phi_1)$, then $t \rightarrow
(x(t,\phi_0)+y(t,\phi_1))$ is decreasing (increasing) on
$[0,t_{L}]$ ($[t_L,\infty)$.)

Now let us consider a point $(\phi_0,\phi_1) \in
\mathbb{B}^{2}_+-\mathbb{C}_0$ such that $\phi_1>-2 \lambda/a$.
Then $t \rightarrow (x(t,\phi_0)+y(t,\phi_1))$ is increasing on
$[0,t_{L}(\phi_0,\phi_1)]$ and is decreasing on
$(t_L(\phi_0,\phi_1),\infty)$ (it decreases to $
-2\lambda/a+\lambda^2/a^2>\lambda/c$). And the monotonicity of $t
\rightarrow x(t,\phi_0)+y(t,\phi_1)$ on $[0,t_{L}(\phi_0,\phi_1)]$
implies that $x(t,\phi_0)+y(t,\phi_1)>\lambda/c$ for $t \in
[0,t_{L}(\phi_0,\phi_1)]$. If $\phi_1=-2\lambda/a$, then
$y(t,\phi_1)=-2 \lambda/a$ for all $t \geq 0$. $x(t,\phi_0)$
increases (decreases) depending on whether $(\phi_0,-2\lambda/a)$
is above or below $L$. These arguments show that if a path
$(x(\cdot,\phi_0), y(\cdot,\phi_1))$ originates in
$\mathbb{B}^{2}_+-\mathbb{C}_0$, then it stays in
$\mathbb{B}^{2}_+-\mathbb{C}_0$. Therefore
 the minimizer $t_{\text{opt}}(\phi_0,\phi_1)$ of
(\ref{detoptstop}) for any $\phi_{0},\phi_{1} \in
\mathbb{B}^{2}_+-\mathbb{C}_{0}$ is equal to zero. Now using
(\ref{ndetopts}) and (\ref{detoptstop}) the optimal continuation
region of (\ref{optstoppingprob}) is equal to $\mathbb{C}$.

\emph{Case II-b-ii-2: $-a/c-1>0.$}  In this case the line segments
$C$ and $L$ intersect at $I=(x^{I},y^{I})$. Arguments similar to
those of Case II-b-i-2 show that
$[(\mathbb{B}^{2}_+-\mathbb{D}_1)\times \mathbb{R}_+] \cap
\mathbb{B}_+^3$, in which $\mathbb{D}_1$ is defined in
(\ref{regd1}), is a superset of the optimal continuation region of
(\ref{optstoppingprob}).
\end{proof}
\begin{thm}\label{thm:sec-5-ikincil}
Let us assume that $\alpha<\beta$ (\emph{Case III:
$\alpha<\beta$}) and define
\begin{equation}\label{eq:defn-D2}
\mathbb{D}_2 \triangleq \left\{(x,y)\in \mathbb{B}^2_{+}: x+y \geq
\frac{\lambda+2 \beta}{c} \right\}.
\end{equation}
Then $ [(\mathbb{B}^{2}_{+}-\mathbb{D}_1) \times \mathbb{R}_+]
\cap \mathbb{B}^3_+$, which is a bounded region in
$\mathbb{R}^3_+$, is an upper bound on the continuation region of
(\ref{optstoppingprob}).
\end{thm}

\begin{proof}
Note that in this case $a>0$. The paths of the processes $\m$,
$\p$ increase between the jumps and decrease with a jump. If $\tau
\in \mathcal{S}$ then there is a constant $t \geq 0$ such that
$\tau \wedge \sigma_1= t \wedge \sigma_1$ almost surely. Hence we
can write
\begin{equation}\label{uppbnd}
\begin{split}
\E_0\left[\int_{0}^{\tau}e^{-\lambda s} h(\Psi_s) ds\right]
&=\E_0\left[\int_0^{\tau \wedge \sigma_1} e^{-\lambda s }h(\Psi_s)
ds\right] +\E_{0}\left[1_{\{\tau \geq
\sigma_1\}}\int_{\sigma_1}^{\tau} e^{-\lambda s}h(\Psi_s)ds\right]
\\&= \E_0\left[\int_0^{t \wedge \sigma_1}
e^{-\lambda s }h(\Psi_s) ds\right]+\E_{0}\left[1_{\{t \geq
\sigma_1\}}\int_{\sigma_1}^{\tau} e^{-\lambda s}h(\Psi_s)ds\right]
\\ & \geq \E_0\left[\int_{0}^{t \wedge \sigma_1}e^{-\lambda s }h(\Psi_s)
ds\right]-\frac{1}{c}\E_{0}\left[1_{\{t \geq \sigma_1\}}
e^{-\lambda \sigma_1}\right]
\\
&=\int_{0}^{t}e^{-(\lambda+2\beta)s}\left[h(x(s,\phi_0),y(s,\phi_1))-\frac{2\beta}{c}\right]ds,
\end{split}
\end{equation}
using also the fact that $\sigma_1$ has exponential distribution
with rate $2 \beta$. From (\ref{uppbnd}) it follows that if
$x(s,\phi_0)+y(s,\phi_1)-(\lambda+2 \beta)/c>0$, then
$\E_0\left[\int_{0}^{\tau}e^{-\lambda s}h(\Psi_s)ds\right]>0$ for
every stopping time $\tau \neq 0$, $\tau \in \mathcal{S}$. Since
the paths $x(t,\phi_0)$, $y(t,\phi_1)$ are increasing we can
conclude that stopping immediately is optimal for
(\ref{optstoppingprob}). That is $\tau=0$ is optimal for
(\ref{optstoppingprob}) if $(\phi_0,\phi_1) \in \mathbb{D}_2$ and
$\phi_2 \leq \phi_1$, in which $\mathbb{D}_2$ is as in
(\ref{eq:defn-D2}).
\end{proof}

Theorems~\ref{thm:sec-5-ilk} and \ref{thm:sec-5-ikincil} can be
used to determine approximate detection rules besides helping us
to determine the location and the shape of the continuation
region. As we have seen in Cases II-b-i-1 and II-b-ii-1, these
approximate rules turn out to be tight. The next theorem is
essential in proving the fact that the smallest optimal stopping
time of (\ref{optstoppingprob}) has a finite expectation.

\begin{thm}\label{thm:finite-expectation}
Let $\tau_D$ be the exit time of the process $\Upsilon$ from a
bounded region $D \subset \mathbb{B}^3_+$. Then
$E_0^{\ms,\ps,\phi^1}[\tau_D]<\infty$ for every
$(\ms,\ps,\phi^1)\in \mathbb{B}^3_+$. Hence $\tau^*$ defined in
(\ref{optst}) has a finite expectation.
\end{thm}
\begin{proof}
Let $f(\ms,\ps,\phi^1) \triangleq \ms+\ps$. Then it follows from
(\ref{eq:infinitesimall}) that
\begin{equation}
\begin{split}
[\mathcal{A} f](\ms,\ps,\phi^1)&=\lambda \ps+ 2 a \ms+ 2 \lambda +
a \ps+ \beta\left[\frac{\alpha}{\beta}\, \ps+ \ms +
\left(\frac{\alpha}{\beta}-1\right)\phi^1-\ms-\ps\right]
\\&\beta\left[\frac{\alpha}{\beta}\, \ps+ \frac{\alpha}{\beta}\,\ms
-
\left(\frac{\alpha}{\beta}-1\right)\phi^1-\ms-\ps\right]=2\lambda(\ms+\ps+1)
\geq 2\lambda,
\end{split}
\end{equation}
for every $(\ms,\ps,\phi^1)\in \mathbb{B}^{3}_+$. Since $f$ is
bounded on $D$, and $\tau_D \wedge t$ is a bounded
$\mathbb{F}$-stopping time, we have
\begin{equation}
\E_0\left[f(\Upsilon_{\tau_D \wedge t})\right]=f(\Upsilon_0)+
\E_0\left[\int_0^{\tau_D \wedge t}[A f](\Upsilon_s)ds\right] \geq
2 \lambda E_0[\tau_D \wedge t].
\end{equation}
On the other hand
\[
\E_0\left[f(\Upsilon_{\tau_D \wedge t})\right]\leq
\frac{\alpha}{\beta}\, \xi,
\]
in which $\xi=\min\{a \in \mathbb{R}_+: \text{for any $(x,y,z) \in
D, \max(x,y,z) \leq a$}\}<\infty$. An application of the monotone
convergence theorem implies that $E_0[\tau_D]<\infty$.
\end{proof}

The results of this section can be used to determine approximate
detection rules besides helping us to determine the location and
the shape of the continuation region. As we have seen  in Cases
II-b-i-1 and II-b-ii-1, these approximate rules turn out to be
tight.

\section{Optimal Stopping with Time Horizon
$\sigma_n$}\label{vnsect} In this section, we will first
approximate the optimal stopping problem (\ref{optstoppingprob})
by a sequence of optimal stopping problems. Let us denote
\begin{equation}\label{defnofVn}
V_{n}(a,b,c) \triangleq \inf_{\tau \in \mathcal{S}}\E^{a,b,c}_0
\left[\int_0^{\tau \wedge \sigma_n }e^{-\lambda
t}h\left(\m_t,\p_t\right)dt \right],
\end{equation}
for all $(a,b,c) \in \mathbb{B}^3_+$ and $n \in \mathbb{N}$. Here,
 $\sigma_n$ is the $n^{\text{th}}$ jump time of the process
$X$.

Observe that $(V_n)_{n \in \mathbb{N}}$ is a decreasing sequence
and they each of its members satisfy $-1/c<V_{n}<0$. Therefore the
pointwise limit  $\lim_n V_n$ exist. It can be shown that more is
true using the fact that the function $h$ is bounded from below
and $\sigma_n$ is a sum of independent exponential random
variables.
\begin{lemma}\label{expofast}
For any $(a,b,c) \in \mathbb{B}^{3}_{+}$
\begin{equation}
\begin{split}
0 \leq V_{n}(a,b,c)-V(a,b,c) & \leq
\frac{1}{c}\left(\frac{2\beta}{2\beta+\lambda}\right)^n.
\end{split}
\end{equation}
\end{lemma}

\begin{proof}
For any $\tau \in \mathcal{S}$,
\begin{equation}\label{bndd}
\begin{split}
 \E_0 \left[\int_0^{\tau}e^{-\lambda
t}h\left(\m_t,\p_t\right)dt \right] &=\E_0 \left[\int_0^{\tau
\wedge \sigma_n }e^{-\lambda t}h\left(\m_t,\p_t\right)dt \right] +
\E_0\left[1_{\{\tau \geq \sigma_n\}} \int_{\sigma_n}^{\tau}
e^{-\lambda t} h(\m_t,\p_t)dt\right]
\end{split}
\end{equation}
The first term on the right-hand-side of (\ref{bndd}) is greater
than $V_n$. Since $h(\cdot,\cdot)>-\lambda/c$ we can show that the
second term is greater than
\begin{equation}
-\frac{\lambda}{c}\E^{\phi_0,\phi_1}_0\left[1_{\{\tau \geq
\sigma_n\}}\int_{\sigma_n}^{\tau}e^{-\lambda s}ds\right] \geq
-\frac{1}{c} \E^{\phi_0,\phi_1}_0\left[e^{-\lambda
\sigma_n}\right] \geq -\frac{1}{c}\left(\frac{2 \beta}{\lambda+
2\beta }\right)^n.
\end{equation}
To show the last inequality we have used the fact that $\sigma_n$
is a sum of $n$ independent and identically distributed
exponential random variables with rate $2 \beta$. Now, the proof
of the lemma follows immediately.
\end{proof}

As in \cite{bdk05} and  \cite{MR96b:90002} to calculate the value
functions $V_n$ iteratively we introduce the
 functional operators $J$, $J_t$. These operators are defined through their
actions on bounded functions $g:\mathbb{B}^{3}_{+} \rightarrow
\mathbb{R}$ as follows:
\begin{equation}\label{defnJ}
\begin{split}
[J g](t,a,b,c) & \triangleq \E^{a,b,c}_{0}\bigg[\int_0^{t \wedge
\sigma_1}e^{-\lambda s} h(\m_s,\p_s)ds +1_{\{t \geq
\sigma_1\}}e^{-\lambda \sigma_1}
g(\m_{\sigma_1},\p_{\sigma_1},\Phi_{\sigma_1}^1) \bigg], \quad
\text{and},
\\ [J_{t}g](a, b,c) & \triangleq \inf_{s \in [t,\infty]}
[J g](s,a, b, c), \quad t \in [0,\infty].
\end{split}
\end{equation}

Observe that
\begin{equation}\label{eq:second-part-of-Jgn}
\begin{split}
&\E_0\left[1_{\{t \geq \sigma_1\}}e^{-\lambda \sigma_1}
g(\m_{\sigma_1},\p_{\sigma_1},\Phi^1_{\sigma_1}) \right]
=\E_0\bigg[\bigg(g\left(\frac{\alpha}{\beta}\m_{\sigma_1-},\p_{\sigma_1-}+\left(\frac{\alpha}{\beta}-1\right)
\Phi^1_{\sigma_1-},
\frac{\alpha}{\beta}\Phi^1_{\sigma_1-}\right)1_{\{X^1_{\sigma_1}
\neq X^1_{\sigma_1-}
\}}\\&+g\left(\frac{\alpha}{\beta}\m_{\sigma_1-},\frac{\alpha}{\beta}\p_{\sigma_1-}-\left(\frac{\alpha}{\beta}-1\right)\Phi^1_{\sigma_1-},
\Phi^1_{\sigma_1-}\right)1_{\{X^2_{\sigma_1} \neq X^2_{\sigma_1-}
\}}\bigg) 1_{\{t \geq \sigma_1\}}e^{-\lambda \sigma_1}\bigg]
\\&=\frac{1}{2}\int_0^{t} 2 \beta e^{-(\lambda+2 \beta)s}g\left(\frac{\alpha}{\beta}\,x(s,a), y(s,b)+\left(\frac{\alpha}{\beta}
-1\right)z(s,c), \frac{\alpha}{\beta}\, z(s,c)\right)ds
\\&+\frac{1}{2}\int_0^{t} 2 \beta e^{-(\lambda+2 \beta)s}g\left(\frac{\alpha}{\beta}\,x(s,a),
\frac{\alpha}{\beta}\,y(s,b)-\left(\frac{\alpha}{\beta}
-1\right)z(s,c), z(s,c)\right)ds.
\end{split}
\end{equation}
To derive (\ref{eq:second-part-of-Jgn}) we used the fact that
 $\sigma_1$ has exponential distribution with
rate $2 \beta$, the dynamics in (\ref{dynpm}), and the fact that
conditioned on the event that there is a jump it has $1/2$
probability of coming from $X^1$ and

Using (\ref{eq:second-part-of-Jgn}) and Fubini's theorem we can
write
\begin{equation}\label{repJ-without-tilde}
[J g](t, a, b, c)= \int_0^{t}e^{-(\lambda+2 \beta)s} (h+  \beta
\cdot g \circ (F_1+F_2))(x(s,a),y(s,b), z(s,c))ds,
\end{equation}
where
\begin{equation}
F_i(a,b,c)=\left(\frac{\alpha}{\beta}\, a,
\left(\frac{\alpha}{\beta}\right)^{i-1} b+(-1)^i
\left(\frac{\alpha}{\beta}-1\right)c,\left(\frac{\alpha}{\beta}\right)^{2-i}\,
c\right) \quad i \in\{1,2\}.
\end{equation}
Using (\ref{detversion}) it can be shown that
\begin{equation} \label{Jfatinfty}
\lim_{t \rightarrow \infty}[J g] (t,a,b,c)=[J
g](\infty,a,b,c)<\infty.
\end{equation}

\begin{lemma}\label{proofJ}
For every bounded function $f$, the mapping $J_0 f$ is bounded. If
$f$ is a concave function, then $J_0 f$ is also a concave
function. If $f_1 \leq f_2$, then $J_0 f_1 \leq J_0 f_2$.
\end{lemma}
\begin{proof}
The third assertion of the lemma directly follows from the
representation (\ref{repJ-without-tilde}). The first assertion
holds since $h$ is bounded from below and $J_{0}f (a,b,) \leq J
f(0,a,b,c)=0$. The second assertion follows from the linearity of
the functions $x(t,\cdot)$, $y(t,\cdot)$, $h(\cdot,\cdot)$,
$F_1(\cdot,\cdot,\cdot)$ and $F_2(\cdot,\cdot,\cdot)$.
\end{proof}

Using Lemma~\ref{proofJ} we can prove the following corollary:
\begin{cor}\label{defnofvn}
Let us define a sequences of function $(v_n)_{n \in \mathbb{N}}$
by
\begin{equation}\label{definevn}
v_0 \triangleq 0, \quad v_n \triangleq J_0 v_{n-1}.
\end{equation}
  Then every $n \in
\mathbb{N}$, $v_n$ is bounded and concave; and $v_{n+1} \leq v_n$.
Therefore, the limit $v=\lim_n v_n$ exist, and is bounded and
concave. Moreover, $v_n$, for all $n \in \mathbb{N}$ and  $v$ are
increasing in each of their arguments.
\end{cor}

\begin{proof}
The proof of the first part directly follows from
Lemma~\ref{proofJ}. That $v_n$ for all $n \in \mathbb{N}$ and
 $v$  are increasing in each of their
arguments follows from the fact that these functions are bounded
from above and below and that they are concave.
\end{proof}

We will need the following lemma  to give a characterization of
the stopping times of the filtration $\mathbb{F}$ (see
\cite{MR82m:60058}).
\begin{lemma}\label{reptau}
For every $\tau \in \mathcal{S}$, there are
$\mathcal{F}_{\sigma_n}$ measurable random variables $\xi_n:
\Omega \rightarrow \infty$ such that $\tau \wedge
\sigma_{n+1}=(\sigma_n+\xi_n) \wedge \sigma_{n+1}$ $\P_0$ almost
surely on $\{\tau \geq \sigma_n\}$.
\end{lemma}
The main theorem of this section can be proven by induction using
Lemma ~\ref{reptau} and the strong Markov property.
\\
\begin{thm}\label{vneVn}
For every $n \in \mathbb{N}$, $v_n$ defined in
Corollary~\ref{defnofvn} is equal to $V_n$. For $\varepsilon \geq
0$, let us denote
\begin{equation}\label{rneps}
\begin{split}
r^{\eps}_{n}(a,b,c) &\triangleq \inf \{t\in (0,\infty]: [J v_n]
(t,(a,b,c)) \leq [J_0 v_n](a,b,c)+\varepsilon \}.
\end{split}
\end{equation}
And let us define a sequence of stopping times by $S^{\eps}_1
\triangleq r_{0}^{\eps}(\Upsilon_0) \wedge \sigma_1$ and
\begin{equation}\label{Sne-without-tilde}
S^{\eps}_{n+1} \triangleq
\begin{cases}
r^{\eps/2}_n(\Upsilon_0) & \text{if $\sigma_1 \geq
r^{\eps/2}_{n}(\Upsilon_0)$}
\\ \sigma_1+S_n^{\eps/2} \circ \theta_{\sigma_1}&
\text{otherwise}.
\end{cases}
\end{equation}

Here $\theta_s$ is the shift operator on $\Omega$, i.e., $X_{t}
\circ \theta_s=X_{s+t}$. Then $S^{\eps}_n$ is an $\eps$-optimal
stopping time of (\ref{defnofVn}), i.e.,
\begin{equation}\label{vnSne}
\E_0^{a,b,c}\left[\int_0^{S^{\eps}_n}e^{-\lambda t}h(\Psi_t)dt
\right] \leq v_{n}(a,b,c)+\eps,
\end{equation}
in which $\Psi_t= (\m_t,\p_t)$, $t \geq 0$.
\end{thm}
\begin{proof}
See Appendix.
\end{proof}

Theorem~\ref{vneVn} shows that the value function $V_n$ of the
optimal stopping problem defined in (\ref{defnofVn}) and the
function $v_n$ introduced in Corollary~\ref{defnofvn} by an
iterative application of the operator $J_0$ are equal. This
implies that the value function of the optimal stopping problem of
(\ref{defnofVn}) can be found by solving a sequence of
deterministic minimization problems.

\section{Optimal Stopping Time}\label{optstotimesect}
\begin{thm}\label{conoptst}
$\tau^*$ defined in (\ref{optst}) is the smallest optimal stopping
time for (\ref{optstoppingprob}).
\end{thm}

This theorem shows that $\Gamma$ defined in (\ref{eq:srop-regn})
is indeed an optimal stopping region. We will divide the proof of
this theorem into several lemmas.

The following dynamic programming principle can be proven by the
special representation of the stopping times of a jump process
(Lemma~\ref{reptau}) and the strong Markov property.
\begin{lemma}\label{dypp}
For any bounded function $g:\mathbb{B}^3_+ \rightarrow \mathbb{R}$
we have
\begin{equation}
[J_t g](a,b,c)=[J g] (t,a, b,c)+e^{-(\lambda+2 \beta)t}[J_0
g](x(t,a),y(t,b),z(t,c)).
\end{equation}
\end{lemma}
Let us denote
\begin{equation}\label{defnrn}
r_n(a,b,c) \triangleq r^{0}_n(a,b,c),
\end{equation}
which is well defined because of (\ref{Jfatinfty}) and the
continuity of the function $t \rightarrow [J f](t,a,b,c)$, $t \geq
0 $. (See (\ref{rneps}) for the definition of $r^{0}_n$.) Let us
also denote
\begin{equation}\label{defnr}
r(a,b,c) \triangleq \inf\{ t \geq 0: [J V](t,a,b,c) =J_0 V(a,b,c)
\}.
\end{equation}
\begin{cor}\label{rnlemm}
The functions $r_{n}$ and $r$ defined by (\ref{defnrn}) and
(\ref{defnr}) respectively, satisfy
\begin{equation}\label{eqdefn}
\begin{split}
& r_n(a,b,c) = \{t \geq 0: v_{n+1}(x(t,a),y(t,b),z(t,c))=0\}
\\ & r(a,b,c) = \{t \geq 0:
v(x(t,a),y(t,b),z(t,c))=0\},
\end{split}
\end{equation}
with the convention that $inf\, \emptyset=0$. Together with
(\ref{eqdefn}), Corollary~\ref{defnofvn} implies that $r_n(a,b,c)
\uparrow r(a,b,c)$ as $n \uparrow \infty$.
\end{cor}
\begin{proof}

Suppose that $r_n(a,b,c)<\infty$. Then from (\ref{rneps}) it
follows that
\begin{equation}\label{eq:Jvnrnabc}
\begin{split}
& [J v_n](r_n(a,b,c),a,b,c)=[J_0
v_n](a,b,c)=[J_{r_n(a,b,c)}v_n](a,b,c)
\\ &=[J v_n](r_n(a,b,c),a,b,c)+e^{-(\lambda+2
\beta)r_n(a,b,c)}v_{n+1}(x(r_n(a,b,c),a),y(r_n(a,b,c),b),z(r_n(a,b,c),c)).
\end{split}
\end{equation}
Here the second equality follows from the definition of the
operator $J_t$ in (\ref{defnJ}) and the third equation follows
from Lemma~ \ref{dypp} and the fact that $J_0v_n=v_{n+1}$. From
(\ref{eq:Jvnrnabc}) it follows that $v_{n+1}(x(r_n(a,b,c),a),$
$y(r_n(a,b,c),b), z(r_n(a,b,c),c))=0$. For $t \in (0,r_n(a,b,c))$
we have $[J v_n](t,a,b,c)$ $>[J_0 v_n](a,b,c)$
$=[J_{r_n(a,b,c)}v_{n}](a,b,c)=J_{t}v_n(a,b,c)$, since the
function $s \rightarrow [J_{s}v_n](a,b,c)$ is non-decreasing.
Using Lemma~ \ref{dypp} we can write
\begin{equation}
\begin{split}
[J_{0}v_n](a,b,c)&=[J_{t}v_n](a,b,c) =
[Jv_n](t,a,b,c)+e^{-(\lambda+2
\beta)}v_{n+1}(x(t,a),y(t,b),z(t,c)),
\end{split}
\end{equation}
which implies that $v_{n+1}(x(t,a),y(t,b),z(t,c))<0$ for all
$t<r_n(a,b,c)$.

Now, suppose that $r_{n}(a,b,c)=\infty$. Then
$v_{n+1}(x(t,a),y(t,b),z(t,c))<0$ for every $t \in (0,\infty)$
which can be shown using same arguments as above. Therefore
$\{t>0: v_{n+1}(x(t,a),y(t,b),z(t,c))=0\}=\emptyset$ and
(\ref{eqdefn}) holds.

The proof for the representation of $r$ can be proven using the
same line of argument and the fact that $J_0 V=V$. The fact that
$J_0 V=V$ can be proven by the dominated convergence theorem,
since the sequences $(v_n(a,b,c))_{n \geq 0}$ and $([J
v_n](t,a,b,c))_{n \geq 0}$ are decreasing, and since $v_n$ is
bounded function for all $n \in \mathbb{N}$.
\end{proof}

In the next lemma we construct optimal stopping times for the
family of problems introduced in (\ref{defnofVn}).
\begin{lemma}
Let us denote $S_n \triangleq S^{0}_n$, where $S^{\eps}_n$ is
defined in Theorem~\ref{vneVn} for $\eps \geq 0$. Then the
sequence $(S_n)_{n \in \mathbb{N}}$ is an almost surely increasing
sequence. Moreover $S_n < \tau^*$ almost surely for all $n$.
\end{lemma}
\begin{proof}
Since $r_1 > 0$, using Corollary~\ref{rnlemm} we can write
\begin{align}
\begin{gathered}
S_{2}-S_{1}=
  \left\{
  \begin{aligned}
    &r_1-r_{0}, && \quad \text{if}\;\; \sigma_1>
      r_1 \\
    &\sigma_1-r_{0} + S_1 \circ \theta_{\sigma_1}, &&
    \quad \text{if}\;\; r_{0}<\sigma_1 \leq r_{1} \\
    & S_{1}\circ \theta_{\sigma_{1}} &&
    \quad \text{if}\;\; \sigma_{1} \leq r_{0}
  \end{aligned}
  \right\} > 0.
  \end{gathered}
  \end{align}
Now, let us assume that $S_n-S_{n-1} > 0$ almost surely. From
Lemma ~\ref{rnlemm} we have that $r_n>r_{n-1}$. Using this fact
and the induction hypothesis we can write

\begin{align}
\begin{gathered}
S_{n+1}-S_{n}=
  \left\{
  \begin{aligned}
    &r_n-r_{n-1}, && \quad \text{if}\;\; \sigma_1>
      r_n \\
    &\sigma_1-r_{n-1} + S_{n} \circ \theta_{\sigma_1}, &&
    \quad \text{if}\;\; r_{n-1}<\sigma_1 \leq r_{n} \\
    & (S_{n}-S_{n-1})\circ \theta_{\sigma_{1}} &&
    \quad \text{if}\;\; \sigma_{1} \leq r_{n-1}
  \end{aligned}
  \right\} > 0,
  \end{gathered}
  \end{align}
which proves the first assertion of the lemma.

From Corollary~\ref{rnlemm} and the definition of $\tau^*$ it
follows that $\tau^{*} \wedge \sigma_1=r \wedge \sigma_1$.
Therefore $\tau^* \wedge \sigma_1
>r_0 \wedge \sigma_1=S_1$, since $r_0<r$. Now, we will assume that
$S_n < \tau^{*}$ and show that $S_{n+1} <\tau^*$. On $\{\sigma_1
\leq r_n\}$ we have that
\begin{equation}
S_{n+1}=\sigma_1+S_{n} \circ \theta_{\sigma_1} < \sigma_1+\tau^{*}
\circ \theta_{\sigma_1}.
\end{equation}
Since $\tau^* \land \sigma_{1}=r \land \sigma_{1}$ and $r>r_n$, if
$\sigma_{1} \leq r_{n}$, then $\tau^* \land
\sigma_{1}=\sigma_{1}$.  Because $\tau^*$ is a hitting time, on
the set $\{\sigma_{1} \leq r_{n}\} \subset \{\sigma_{1} \leq
\tau^*\}$ the following holds
\[
S_{n+1}\leq \sigma_{1}+\tau^* \circ \theta_{\sigma_{1}}=\tau^*.
\]
On the other hand if $\sigma_{1}>r_{n}$, then $\tau^* \land
\sigma_{1}=r \land \sigma_{1} > r_{n}$. Therefore on
$\{\sigma_{1}>r_{n}\}$, $S_{n+1}=r_{n} < \tau^*$. This concludes
the proof of the second assertion.
\end{proof}
\begin{lemma}\label{defns0}
Let us denote $\Psi_t=(\m_t,\p_t)$, $t \geq 0$. If $S^* \triangleq
\lim_{n}S_{n}$, then $S^* = \tau^*$ almost surely. Moreover,
$\tau^*$ is an optimal stopping time, i.e.,
\[
V(a,b,c)=\E^{a,b,c}_0 \left[\int^{S^*}_0 e^{-\lambda s}
    h(\Psi_s)ds \right].
\]
\end{lemma}

\begin{proof}
The limit $S^* \triangleq \lim_{n}S_{n}$ exists since $(S_n)_{n
\in \mathbb{N}}$ is increasing and $S_n \leq \tau^*<\infty$ (as a
corollary of Theorem~\ref{thm:finite-expectation}) for all $n$.
Let us show that $S^*$ is optimal.
\begin{equation}\label{fatoulem}
\begin{split}
\E_{0}\left[\lim_{n}\int_{0}^{S_{n}}e^{-\lambda t} h(\Psi_{t})dt
\right]&\leq \liminf_{n} \E_{0}\left[\int_{0}^{S_{n}}e^{-\lambda
t} h(\Psi_t)dt \right]
\\ &=\lim_{n}
V_{n}(a,b,c)=V(a,b,c).
\end{split}
\end{equation}
The first inequality follows from Fatou's Lemma, which we can
apply since
\[
\int_{0}^{S_{n}}e^{-\lambda t}h(\Psi_t)dt \geq
\int_{0}^{\infty}e^{-\lambda t}h(\Psi_t)dt \geq
-\frac{\sqrt{2}}{c}, \ \ \text{a.s.}
\]
The first equality in (\ref{fatoulem}) follows from
Theorem~\ref{vneVn}. Now it can be seen from (\ref{fatoulem}) that
$S^*$ is an optimal stopping time. Taking the limit of
(\ref{Sne-without-tilde}) as $n \rightarrow \infty$ and using
Corollary~\ref{rnlemm} we conclude that $\tau^*=S^*$.
\end{proof}

\emph{Proof of Theorem~\ref{conoptst}} The proof of the optimality
of $\tau^*$ follows directly from Lemma~\ref{defns0}. We will show
that $\tau^*$ is the smallest optimal stopping time.

Given any $\mathbb{F}$-stopping time $\tau<\tau^*$, let us define
\begin{equation}
\tilde{\tau}\triangleq
 \tau+\tau^* \circ \theta_{\tau}
\end{equation}
Then the stopping time $\tilde{\tau}$ satisfies
\begin{equation}\label{strngmarkv}
\begin{split}
\E_{0}^{a,b,c}\left[\int_0^{\tilde{\tau}}e^{-\lambda s}
h(\Psi_s)ds \right]
 &=\E_0^{a,b,c}\left[\int_0^{\tau}e^{-\lambda s}h(\Psi_s)ds+\int_{\tau}^{\tilde{\tau}}e^{-\lambda s} h(\Psi_s)ds\right]
\\&=\E_0^{a,b,c}\left[\int_0^{\tau}e^{-\lambda s}h(\Psi_s)ds+e^{-\lambda \tau} \int_0^{\tau^*\circ \theta_{\tau}}
e^{-\lambda s} h(\Psi_{s+\tau})ds\right]
\\ &=\E_0^{a,b,c}\left[\int_0^{\tau}e^{-\lambda s}h(\Psi_s)ds+  e^{-\lambda \tau}V(\Upsilon_{\tau})\right]
\\ &<
\E_0^{a,b,c}\left[\int_0^{\tau}e^{-\lambda s}h(\Psi_s)ds\right].
\end{split}
\end{equation}
Here the third equality follows from the strong Markov property of
the process $\Upsilon$ and the inequality follows since
$V(\Upsilon_{\tau})< 0$. Equation (\ref{strngmarkv}) shows that
any optimal stopping time $\tau<\tau^*$ cannot be optimal.

\hfill $\square$

\section{Structure of the Continuation and Stopping Regions
Regions}\label{stofopt}

Let us recall (\ref{eq:srop-regn}) and denote
\begin{equation}\label{optregn}
\begin{split}
\mathbf{\Gamma_n} &\triangleq \{(a,b,c) \in \mathbb{B}^{3}_{+}:
v_n(a,b,c)=0\}, \quad \mathbf{C}_n \triangleq
\mathbb{B}^{3}_{+}-\mathbf{\Gamma_n}.
\end{split}
\end{equation}

We have shown in Theorem~\ref{conoptst} that that $\Gamma$ of
(\ref{eq:srop-regn}) is the optimal stopping region for
(\ref{optstoppingprob}) and the first hitting time $\tau^*$ of
$\Upsilon$ to this set is optimal. On the other hand although
$\Gamma_n$ is an optimal stopping region for (\ref{defnofVn}), the
description of the optimal stopping times
 $S_{n}^{0}$ (see (\ref{Sne-without-tilde})) is more involved. These optimal
stopping times are not hitting times of the sets $\Gamma_n$.
$S^0_n$ prescribes to stop if $\Upsilon$ hits $\Gamma_n$ before it
jumps. Otherwise if there is a jump before $\Upsilon$ reaches
$\Gamma_n$, then $S^0_n$ prescribes to stop when the process hits
$\Gamma_{n-1}$ before the next jump, and so on.

Theorem \ref{vneVn} shows that $V_{n}$ of (\ref{defnofVn}) and the
functions $v_n$ introduced in Corollary \ref{defnofvn} are equal.
Therefore, their respective limits $V$ and $v$ are also equal.
Recall that $V^n$ converges to $V$ uniformly and the convergence
rate is exponential (see Lemma \ref{expofast}). Since $(v_{n})_{n
\in \mathbb{N}}$ is a decreasing sequence with limit $v$ the
stopping regions in (\ref{optregn}) are nested and satisfy $\Gamma
\subset \cdots \subset \Gamma_n \subset \Gamma_{n-1} \subset
\cdots \Gamma_1$ and $\Gamma= \cap_{n=1}^{\infty} \Gamma_n$.

By Corollary \ref{defnofvn} we know that each $v_n$ is concave and
bounded, which also implies that the limit $v$ is concave and
bounded. This in turn implies that the stopping regions $\Gamma_n$
and $\Gamma$ are convex and closed. Since we show in Section
\ref{reginbnd} that the continuation region is bounded, it can
readily be shown that the stopping regions $\Gamma_n$ and $\Gamma$
are the epigraphs of some mappings $\gamma_n$ and $\gamma$ which
are convex and strictly decreasing and the numbers $x_n \triangleq
\inf\{y \in\mathbb{R_+}:\gamma_n(y)=0\}$ and $x \triangleq \inf\{y
\in \mathbb{R_+}: \gamma(x)=0\}$ are finite.
%NOTE (08.08.05):
%The concavity and convexity arguments are not true. Therefore it seems like the
%optimal stopping time is not an exit time from a convex region.

\section{Extensions}\label{sec:extensions}

\subsection{Non-identical Sources}
Consider two independent Poisson processes $X^{1}$ and $X^2$ with
arrival rates $\beta_1$ and $\beta_2$ respectively. At some random
unobservable times $\theta_1$ and $\theta_2$, with distributions
\begin{equation}\label{edist-2}
\P(\theta_i=0)=\pi_i \quad \P(\theta_i>t)=(1-\pi_i)e^{-\lambda_i
t}\,\, \text{for}\,\, t \geq 0,
\end{equation}
the arrival rates of the Poisson processes $X^1$ and $X^2$ change
from $\beta_i$ to $\alpha_i$, respectively, i.e.,
\begin{equation}\label{intensita-2}
X^i_{t}-\int_{0}^{t}h_i(s)ds, \quad t \geq 0,\,i=1,2,
\end{equation}
are martingales, in which
\begin{equation}
h_i(t)=[\beta_i 1_{\{s<\theta_i\}}+\alpha_i 1_{\{s \geq
\theta_i\}}], \quad t \geq 0, \, i=1,2.
\end{equation}
 Here $\alpha_1$, $\alpha_2$, $\beta_1$ and $\beta_2$ are known
positive constants. Then Equation the dynamics of $\m$ defined in
(\ref{eq:defn-m-p-5}) becomes
\begin{equation}\label{dynpm-2}
\begin{split}
d \m_t&= [\lambda_2 \Phi^1_t+ \lambda_1
\Phi^2_t+(a_1+a_2)\p_t]dt+\m_t \left[((\alpha_1/\beta_1)-1)
dX^1_t+((\alpha_2/\beta_2)-1)d X^2_t\right],
\end{split}
\end{equation}
in which $a_i=\lambda_i-\alpha_i+\beta_i$, $i \in \{1,2\}$. Let us
introduce
\begin{equation}\label{detversion-2}
\begin{split}
x(t,\phi_0)&=e^{(a_1+a_2)t} \phi_0+\int_0^{t}e^{(a_1+a_2)(t-u)}
(\lambda_2 y(u,\phi_1)+\lambda_1 z(u,\phi_2))du, \quad \text{in
which}
\\ y(t,\phi_1)&=-\frac{\lambda_1}{a_1}+e^{a_1t}\left(\phi_1+\frac{
\lambda}{a}\right), \quad z(t,\phi_2)=-\frac{
\lambda_2}{a_2}+e^{a_2 t}\left(\phi_2+\frac{
\lambda_2}{a_2}\right).
\end{split}
\end{equation}
Then $\m_t$, $\Phi^1_t$ and $\Phi^2_t$, $t \geq 0$ can be written
as
\begin{equation}
\m_t=x(t-\sigma_n,\m_{\sigma_n}), \,\,
\Phi^1_t=y(t-\sigma_n,\p_{\sigma_n}),\,\,
\Phi^2_t=z(t-\sigma_n,\Phi^1_{\sigma_n}) \quad \sigma_n \leq
t<\sigma_{n+1}, \, n\in \mathbb{N},
\end{equation}
and {\small
\begin{equation}
\begin{split}
 \m_{\sigma_{n+1}}&=\left(\frac{\alpha_1}{\beta_1}1_{\{X^{1}_{\sigma_{n+1}}\neq X^{1}_{\sigma_{n+1-}
}\}}+\frac{\alpha_2}{\beta_2}1_{\{X^{2}_{\sigma_{n+1}}\neq
X^{2}_{\sigma_{n+1-}}\}}
 \right)\,\m_{\sigma_{n+1-}},
\\ \Phi^1_{\sigma_{n+1}}&=\frac{\alpha_1}{\beta_1}1_{\{X^{1}_{\sigma_{n+1}}\neq
X^{1}_{\sigma_{n+1-} }\}}\,\Phi^1_{\sigma_{(n+1)-}}, \quad
\Phi^2_{\sigma_{n+1}}=\frac{\alpha_2}{\beta_2}1_{\{X^{2}_{\sigma_{n+1}}\neq
X^{2}_{\sigma_{n+1-} }\}}\,\Phi^1_{\sigma_{(n+1)-}}.
\end{split}
\end{equation}
Choosing $\Upsilon_t=\left(\m_t,\Phi^1_t,\Phi^2_t\right)$, $t \geq
0$ as the Markovian statistic to work with, we can extend our
analysis to deal with non-identical sources.

\subsection{When there are more than two sources}
We have solved a two-source quickest detection problem in which
the aim is to detect the minimum of two disorder times. Our
approach can easily be generalized to problems including several
dimensions. To clarify how this generalization works, let us show
what the sufficient statistics are when there are three
independent sources. Assume that the observations come from the
independent sources $X^1$, $X^2$ and $X^3$. Let $\Phi_t$ be the
odds ratio defined in (\ref{eq:defn-phi-t}). Then
\begin{equation}
\Phi_t=\Phi^1_t+\Phi^2_t+\Phi^3_t+\Phi^1_t\Phi^2_t+\Phi^1_t\Phi^3_t+\Phi^2_t\Phi^3_t+\Phi^1_t\Phi^2_t\Phi^3_t,
\end{equation}
in which $\Phi^i$, $i\in\{1,2,3\}$ is defined as in
(\ref{eq:phi-t-i}). Let us denote $\Phi_t^{(i,j)}\triangleq
\Phi^{i}_t \Phi^j_t$, $i,j \in \{1,2,3\}$ and
$\Phi^{(x)}_t\triangleq \Phi^1_t\Phi^2_t\Phi^3_t$, $t\geq 0$. The
dynamics of these processes can be written as
\begin{equation}\label{eq:dyn-gen-multi}
\begin{split}
d\Phi^{(i,j)}_t&=[\lambda(\Phi^i_t+\Phi^j_t)+2(\lambda-\alpha+\beta)\Phi^{(i,j)}_t]dt+
\left(\frac{\alpha}{\beta}-1\right)\Phi^{(i,j)}_t d(X_t^i+X^j_t),
\\ d\Phi^{(x)}_t
&=\left[\lambda\left(\Phi^{(1,2)}_t+\Phi^{(1,3)}_t+\Phi^{(2,3)}_t\right)+3(\lambda-\alpha+\beta)\Phi^{(x)}_t\right]dt
+\left(\frac{\alpha}{\beta}-1\right)\Phi^{(x)}_t
d(X^1_t+X^2_t+X^3_t).
\end{split}
\end{equation}

We can see from (\ref{penalty}) and (\ref{eq:dyn-gen-multi}) that
$\Upsilon\triangleq(\Phi^1, \Phi^2, \Phi^3, \Phi^{(1,2)}_t,
\Phi^{(1,3)}_t, \Phi^{(2,3)}_t, \Phi^{(x)})$ is a 7 dimensional
Markovian statistic whose natural filtration is equal to the
filtration generated by $X^1$, $X^2$ and $X^3$. From this one can
see that the results of Sections \ref{vnsect} and
\ref{optstotimesect} can be extended to the three dimensional case
since these results rely only on the fact that the sufficient
statistic $\Upsilon$ is a strong Markov process. The boundedness
of the continuation region can also be shown as in
Section~\ref{reginbnd} since these results can be derived from the
sample path properties of the sufficient statistic.

As a result, our results are applicable for decision making with
large-scale distributed networks of information sources. In the
future, using the techniques developed here, we would like to
solve a multi-source detection problem where the observations come
from correlated sources. We also would like to extend our results
and develop change detection algorithms that can be applied
effectively to multiple source data that involves both continuous
and discrete event phenomena.

\subsection{When the jump size of the observations are random}

Consider two independent \emph{compound} Poisson processes
$X^{i}=\{X^{i}_t:t \geq 0\}$, $i \in \{1,2\}$, where
\begin{equation}\label{eq:X}
X^{i}_{t}= X^i_{0}+ \sum_{j=1}^{N^i_{t}} Y^i_{j}\, ,
\end{equation}
in which $N^i$, $i \in \{1,2\}$ are two independent Poisson
processes whose common rate $\beta>0$ changes to $\alpha$ at some
random unobservable times $\theta_i$, $i\in \{1,2\}$,
respectively. The random variables $Y^i_j \in \mathbb{R}^d$, $i
\in \{1,2\}$, which are also termed as `marks', are independent
and identically distributed with a common distribution, $\nu$,
which is called as the `mark distribution'. At the change time
$\theta_i$ the mark distribution of the process $X^i$ changes from
$\nu$ to $\mu$.  We will assume that $\mu$ is absolutely
continuous with respect to $\nu$ and denote the Radon-Nikodym
derivative by $ r(y) \triangleq \frac{d \mu}{d \nu} (y), \,\, y
\in \mathbb{R}^d$. In this case $L_t^i$ in (\ref{likelihood})
becomes
\begin{equation}
L^i_{t}=e^{-(\alpha-\beta)t}
\prod_{k=1}^{N^i_{t}}\left(\frac{\alpha}{\beta}\, r(Y^i_k)\right)
\end{equation}
The likelihood ratio process $L^i$ is the unique solution of the
stochastic differential equation (see e.g. \cite{JS})
\begin{equation}
dL^i_{t}=L^{i}_t \left(-(\alpha-\beta)dt + \int_{y \in
\mathbb{R}^d} \left(\frac{\alpha}{\beta}\, r(y)-1\right) p(dt dy
)\right), \quad L^i_0=1,
\end{equation}
where $p$ is a random measure that is defined as
\begin{equation}
p^i((0,t] \times A) \triangleq \sum_{k=1}^{\infty} 1_{\{\sigma^i_k
\leq t\}} 1_{\{Y^i_{k} \in A\}}, \quad t \geq 0,
\end{equation}
and for any $A$ that is a Borel measurable subset of
$\mathbb{R}^d$. Here $\sigma^i_k$ is the $k^{\text{th}}$ jump time
of the process $X^i$. Now using the change of variable formula for
semi-martingales (see e.g. \cite{Protter}), we can write
\begin{equation}
d \Phi^{i}_{t}=(\lambda+(\lambda-\alpha+\beta) \Phi^{i}_{t})dt+
\Phi^i_{t-} \int_{y \in \mathbb{R}^d}\left(\frac{\alpha}{\beta}\,
r(y)-1\right)p^i(dt dy), \quad \Phi_0^i=\frac{\pi_i}{1-\pi_i},
\end{equation}
for $t \geq 0$, and $i \in \{1,2\}$. Note that
$\Phi^{i}_{\sigma_n}=\frac{\alpha}{\beta}\,
r(Y_n)\Phi^{i}_{\sigma_n-}$ at the $n$th jump time of the process
$X^i$. Using a change of variable formula for semi-martingales,
the dynamics of $\m$ and $\p$ in (\ref{eq:defn-m-p-5}) can be
written as
\begin{equation}
\begin{split}
d \m_t&= [\lambda \p_t+a \m_t]dt+ \m_{t-} \int_{y \in
\mathbb{R}^d}\left(\frac{\alpha}{\beta}\, r(y)-1\right) (p^1+p^2)
(dt dy),
\\ d\Phi^{+}_{t}&=[2 \lambda+
a\Phi^{+}_t]dt+\Phi^1_{t-} \int_{y \in
\mathbb{R}^d}\left(\frac{\alpha}{\beta}\, r(y)-1\right)p^1(dt dy)+
\Phi^2_{t-} \int_{y \in \mathbb{R}^d}\left(\frac{\alpha}{\beta}\,
r(y)-1\right)p^2(dt dy)
\end{split}
\end{equation}
with initial conditions $\m_0=\pi_1 \pi_2/[(1-\pi_1)(1-\pi_2)]$,
and $\p_0=\pi_1/(1-\pi_1)+\pi_2/(1-\pi_2)$.

The bounds on the continuation region constructed for the simple
Poisson disorder problem in Section~\ref{reginbnd} can also be
shown to bound the continuation region of the compound Poisson
disorder problem. On the other hand the results in Sections
\ref{vnsect} and \ref{optstotimesect} can be shown to hold. The
only change will be the form of the operator $J$ in
(\ref{repJ-without-tilde}). But this new operator can be shown to
share the same properties as its counterpart for the un-marked
case.

\section*{Acknowledgment}

This work was supported in part by the U.S. Army Pantheon Project
and National Science Foundation under Grant DMS-0604491. We are
grateful to the two referees for their detailed comments that
helped us improve the manuscript. We also would like to thank
Semih Sezer for insightful comments.
\section{Appendix}

\textbf{Proof of Theorem~\ref{vneVn}}.  We will prove only that
$V_n=v_n$ and $S^{\eps}_n$ is an $\eps$-optimal stopping time of
(\ref{defnofVn}).

The proof will be carried out in three steps.
\\(i) First we will show that $V_{n} \geq v_{n}$. To
establish this fact, it is enough to show that
 for any stopping time $\tau \in \mathcal{S}$
\begin{equation}\label{vnlesseq}
\E_0^{a,b,c}\left[\int_{0}^{\tau \wedge \sigma_n}e^{-\lambda t}
h(\Psi_t) dt\right] \geq v_n(a,b,c).
\end{equation}
In order to prove (\ref{vnlesseq}) we will show that
\begin{equation}\label{induk}
\begin{split}
&\E_0\left[\int_{0}^{\tau \wedge \sigma_n}e^{-\lambda t} h(\Psi_t)
dt\right] \geq \E_0\left[\int_{0}^{\tau \wedge
\sigma_{n-k+1}}e^{-\lambda t} h(\Psi_t) dt+1_{\{\tau \geq
\sigma_{n-k+1}\}} e^{-\lambda \sigma_{n-k+1}}
v_{k-1}(\Upsilon_{\sigma_{n-k+1}})\right],
\end{split}
\end{equation}
for $k \in \{1,2,...,n+1\}$. Note that (\ref{vnlesseq}) follows
from (\ref{induk}) if we set $k=n+1$. In what follows we will to
show (\ref{induk}) by induction.

 When $k=1$, (\ref{induk}) is satisfied since
$v_{0}=0$. Assume that (\ref{induk}) holds for $1 \leq k \leq n+1
$. Let us denote the right-hand-side of (\ref{induk}) by
$\rho_{k-1}$. We can write $\rho_{k-1}=\rho^1_{k-1}+\rho^2_{k-1}$,
where
\begin{equation}\label{defnofrhos}
\begin{split}
& \rho^1_{k-1} \triangleq \E_0\left[\int_{0}^{\tau \wedge
\sigma_{n-k}}e^{-\lambda t} h(\Psi_t) dt\right] \quad \text{and}
\\ &\rho^2_{k-1} \triangleq \E_0\left[1_{\{\tau \geq \sigma_{n-k}\}}
\left(\int_{\sigma_{n-k}}^{\tau \wedge \sigma_{n-k+1}}e^{-\lambda
t} h(\Psi_t) dt+ 1_{\{\tau \geq \sigma_{n-k+1}\}}e^{-\lambda
\sigma_{n-k+1}} v_{k-1}(\Upsilon_{\sigma_{n-k+1}})\right)\right].
\end{split}
\end{equation}
Now by Lemma~\ref{reptau}, there exists an
$\mathcal{F}_{\sigma_{n-k}}$-measurable random variable
$\xi_{n-k}$ such that
\begin{equation}\label{represent}
\tau \wedge \sigma_{n-k+1}=(\sigma_{n-k}+\xi_{n-k}) \wedge
\sigma_{n-k+1} \quad \text{almost surely on $\{\tau \geq
\sigma_{n-k}\}$}.
\end{equation}
Equation (\ref{represent}) together with the strong Markov
property of $\Upsilon$ (with respect to the filtration
$\mathbb{F}$) implies that
\begin{equation}
\rho^2_{k-1}=  \E_0\left[1_{\{\tau \geq \sigma_{n-k}\}}e^{-\lambda
\sigma_{n-k} }f_{k-1}(\xi_{n-k},\Upsilon_{\sigma_{n-k}})\right],
\end{equation}
in which
\begin{equation}
\begin{split}
f_{k-1}(r,(a,b,c)) &\triangleq \E_0^{a,b,c}\left[\int_{0}^{r
\wedge \sigma_1}e^{-\lambda t} h(\Psi_t) dt + 1_{\{r \geq
\sigma_1\}}e^{-\lambda
\sigma_1}v_{k-1}(\Upsilon_{\sigma_1})\right]
\\&=Jv_{k-1}(r,(a,b,c)) \geq J_{0}
v_{k-1}(a,b,c)=v_{k}(a,b,c),
\end{split}
\end{equation}
in which the second equality and the first inequality follow from
(\ref{defnJ}) and the last equality follows from (\ref{definevn}).
Therefore
\begin{equation}\label{inrhok1}
\rho^2_{k-1} \geq \E_0\left[1_{\{\tau \geq
\sigma_{n-k}\}}e^{-\lambda \sigma_{n-k} }
v_k(\Upsilon_{\sigma_{n-k}})\right].
\end{equation}
Now using (\ref{induk}), (\ref{defnofrhos}) and (\ref{inrhok1}) we
obtain that (\ref{induk}) holds when $k$ is replaced by $k+1$. At
this point we have proved by induction that (\ref{induk}) holds
for $k=1,2,...,n+1$.

(ii) The converse of (i), $V_{n} \leq v_n$, follows from
(\ref{vnSne}), since $S^{\eps}_{n} \leq \sigma_n$ by construction
(see (\ref{Sne-without-tilde})).

(iii) What is left to prove is (\ref{vnSne}). If $n=1$, then the
left-hand-side of (\ref{vnSne}) becomes
\begin{equation}
\begin{split}
\E_0^{a,b,c}\left[\int_{0}^{r_{0}^{\eps}(a,b,c) \wedge \sigma_1
}e^{-\lambda t} h(\Psi_t)dt\right] &=J
v_0(r^{\eps}_{0}(a,b,c),a,b,c)  \\ &\leq J_0 v_0
(a,b,c)+\eps=v_{1}(a,b,c)+\eps.
\end{split}
\end{equation}
Now, suppose that (\ref{vnSne}) holds for all $\eps>0$ for some
$n$. Using the fact that $S^{\eps}_{n+1} \wedge
\sigma_1=r^{\eps/2}_{n} \wedge \sigma_1$ almost surely and the
strong Markov property of $\Upsilon$, we can write
\begin{equation}\label{snp1}
\begin{split}
\E_0&\left[\int_{0}^{S^{\eps}_{n+1}}e^{-\lambda t}
h(\Psi_t)dt\right]=\E_0\left[\int_{0}^{S^{\eps}_{n+1}\wedge
\sigma_1 }e^{-\lambda t} h(\Psi_t)dt+ 1_{\{S^{\eps}_{n+1}\geq
\sigma_1 \}} \int_{\sigma_1}^{S^{\eps}_{n+1}}e^{-\lambda t}
h(\Psi_t)dt\right]
\\&=\E_0\left[\int_{0}^{r^{\eps/2}_{n}(a,b,c) \wedge \sigma_1}e^{-\lambda t}
h(\Psi_t)dt\right]+ \E_0\left[1_{\{r^{\eps/2}_{n}(a,b,c) \geq
\sigma_1 \}} e^{-\lambda \sigma_1}
g_{n}(\Upsilon_{\sigma_1})\right],
\end{split}
\end{equation}
in which
\begin{equation}\label{gnphi0phi1}
g_{n}(a,b,c) \triangleq
\E_0^{a,b,c}\left[\int_{0}^{S^{\eps/2}_{n}}e^{-\lambda t}
h(\Psi_t)dt\right] \leq v_{n}(a,b,c)+\eps/2.
\end{equation}
The inequality in (\ref{gnphi0phi1}) follows from the induction
hypothesis. Using (\ref{gnphi0phi1}) we can write (\ref{snp1}) as
\begin{equation}
\begin{split}
\E_0^{a,b,c}\left[\int_{0}^{S^{\eps}_{n+1}}e^{-\lambda t}
h(\Psi_t)dt\right] &\leq \E_0^{a,b,c}
\bigg[\int_{0}^{r^{\eps/2}_{n}(a,b,c) \wedge \sigma_1}e^{-\lambda
t} h(\Psi_t)dt+ 1_{\{r^{\eps/2}_{n}(a,b,c) \geq \sigma_1
\}}e^{-\lambda \sigma_1}
v_{n}(\Upsilon_{\sigma_1})\bigg]+\eps/2\\&=Jv_{n}(r^{\eps/2}_{n}(a,b,c),a,b,c)+\eps/2
 \leq v_{n+1}(a,b,c)+\eps.
\end{split}
\end{equation}
This proves (\ref{vnSne}) when $n$ is replaced by $n+1$. \hfill
$\square$

\bibliographystyle{siam}

\end{document}